%% file: Inversion_ring_SSY_arXiv.tex
\documentclass[aps,pre,10pt,showpacs,floatfix,onecolumn,nofootinbib,a4paper]{revtex4-2}
\usepackage{amssymb, amsmath, amsthm, mathtools}
\usepackage{bm}
\usepackage{mathrsfs}
\usepackage{hyperref,bookmark}
\usepackage{graphicx,colortbl}
\usepackage{epsfig}
\usepackage{mathrsfs}
\usepackage{verbatim}
\usepackage{url}
\usepackage{subcaption}
\usepackage{float}
\usepackage[greek,english]{babel}
\usepackage{dcolumn}

\usepackage{color}
\definecolor{lightgray}{gray}{0.8}

\newcommand{\tr}{\operatorname{tr}}

\newcommand{\diver}{\operatorname{div}}
\newcommand{\curl}{\operatorname{curl}}

\newcommand{\n}{\bm{n}}
\newcommand{\e}{\bm{e}}
\newcommand{\x}{\bm{x}}

\newcommand{\WQT}{W_\mathrm{QT}}
\newcommand{\bend}{\bm{b}}

\newcommand{\R}{\mathbf{R}}

\newcommand{\nT}{\n_{\mathrm{T}}}

\newcommand{\temp}{\mathrm{T}}

\begin{document}
\latintext

\title{Inversion Ring in Chromonic Twisted Hedgehogs: Theory and Experiment}
\author{Federica Ciuchi}
\email{federica.ciuchi@cnr.it}
\affiliation{CNR-Nanotec c/o Physics Department, University of Calabria, Ponte Bucci, Cubo 31C, 87036 Arcavacata di Rende, Italy}
\author{Maria Penelope De Santo}
\email{maria.desanto@fis.unical.it}
\affiliation{CNR-Nanotec c/o Physics Department, University of Calabria, Ponte Bucci, Cubo 31C, 87036 Arcavacata di Rende, Italy}
\author{Silvia Paparini}
\email{silvia.paparini@unipv.it}
\affiliation{Department of Mathematics, University of Pavia, Via Ferrata 5, 27100 Pavia, Italy}
\author{Lorenza Spina}
\email{loryspina2@gmail.com}
\affiliation{CNR-Nanotec c/o Physics Department, University of Calabria, Ponte Bucci, Cubo 31C, 87036 Arcavacata di Rende, Italy}
\affiliation{Physics Department, University of Calabria, Ponte Bucci cubo 33B, Arcavacata di Rende 87036, CS, Italy}
\author{Epifanio G. Virga}
\email{eg.virga@unipv.it}
\affiliation{ Department of Mathematics, University of Pavia, Via Ferrata 5, 27100 Pavia, Italy }

\begin{abstract}
\emph{Twisted hedgehogs} are defects in spherical cavities with homeotropic anchoring for the nematic director that arise when twist distortions are sufficiently less energetic than splay (and bend) distortions. They bear a characteristic \emph{inversion ring}, where the director texture changes the sense it spirals about the center of the cavity. This paper applies a quartic twist theory recently proposed to describe the elasticity of chromonics to explain a  series of inversion rings observed in aqueous solutions of SSY at two different concentrations. The theory features a phenomenological length $a$, whose measure is extracted from the data and shown to be fairly independent of the cavity radius, as expected for a material constant.
\end{abstract}

\date{\today}

\maketitle

\section{Introduction}\label{sec:intro}
Among the many lyotropic phases, chromonic liquid crystals (CLCs) have attracted attention for their potential applications in life sciences \cite{shiyanovskii:real-time,mushenheim:dynamic,mushenheim:using,zhou:living}. These materials are constituted by plank-shaped molecules that arrange themselves in stacks when dissolved in (usually aqueous) solution. For sufficiently large concentrations or small temperatures, the constituting stacks give rise to an ordered phase, either nematic or columnar \cite{lydon:chromonic_1998,lydon:handbook,lydon:chromonic_2010,lydon:chromonic,dierking:novel}. Here, we shall only be concerned with the nematic phase. Numerous substances have a CLC phase; these include organic dyes (especially those common in food industry), drugs, and oligonucleotides. 

The classical elastic Oseen-Frank theory has been applied to CLCs, but the twist constant $K_{22}$ has to be taken anomalously small  to explain the spontaneous \emph{double} twist exhibited by these materials when confined to cylindrical cavities with degenerate planar anchoring on the lateral boundary \cite{davidson:chiral,eun:effects}. In particular, $K_{22}$ must be smaller than the saddle-splay constant $K_{24}$, thus violating one of the inequalities Ericksen~\cite{ericksen:inequalities} had put forward to guarantee that the Oseen-Frank stored energy be bounded below. However, as shown in \cite{paparini:paradoxes}, free-boundary problems may reveal paradoxical consequences stemming from violating Ericksen's inequalities.   If $K_{22}<K_{24}$, a CLC droplet in an isotropic fluid enforcing degenerate planar anchoring, is predicted to be unstable against \emph{shape} perturbations: it would split indefinitely in smaller droplets, while the total free energy plummets to negative infinity \cite{paparini:paradoxes}.
This prediction is in sharp contrast with the wealth of experimental observations of CLC tactoids in the biphasic region, where nematic and isotropic phases coexist in equilibrium \cite{tortora:self-assembly,tortora:chiral,peng:chirality,nayani:using,shadpour:amplification}. 

To resolve this contradiction,   a minimalist quartic theory for CLCs was proposed in \cite{paparini:elastic}, which adds to the Oseen-Frank energy density a single quartic term in the twist measure; hence the name \emph{quartic twist} theory. It was  shown in \cite{paparini:elastic} that indeed within this theory the total free energy of chromonic droplets subject to degenerate planar interfacial anchoring remains bounded below, even if $K_{22}<K_{24}$.   

In \cite{paparini:spiralling}, the quartic twist theory was applied to explain the formation of \emph{inversion rings} within spherical cavities enclosing water solutions of SSY (Sunset Yellow, a dye used in industrially processed food) subject to \emph{homeotropic} boundary conditions for the director field $\n$. The predictions of the theory were contrasted with some experimental evidence found in recent literature \cite{spina:intercalation}.

Although the comparison was encouraging, it was neither systematic nor self-consistent: it was based on a single example and used an estimate for a phenomenological length $a$ featuring in the theory that came from a different experiment. Thus, the motivation for the present study is to provide a systematic and self-consistent validation of the quartic twist theory.

We observed  a large number of microcavities enclosing a SSY solution at two distinct concentrations and three different times since the samples were prepared. We extracted the length $a$ in different physical conditions from direct measurements of the inversion ring.

The paper is organized as follows.  In Section~\ref{sec:theory}, we summarize the quartic twist theory to make our development self-contained. In Section~\ref{sec:materials}, we describe the materials and the experimental methods. Section~\ref{sec:comparison} is devoted to the comparison of the experimental data with the proposed theory. Here we extract $a$ and study its dependence on concentration and other factors that may change the physical conditions of the samples as time elapses since their preparation. Finally, in Section~\ref{sec:conclusion}, we collect the conclusions reached in this study. The paper is closed by two Appendices: in one, we disaggregate the data that in the main text were presented in a single universal graph; in the other, we delve in some statistical details that could have hampered our presentation in the body of the paper.

\section{Quartic Twist Theory}\label{sec:theory}
In this study we adopt an elastic theory for CLCs that has already been developed and applied in a number of papers \cite{paparini:elastic,paparini:spiralling,paparini:what_arxiv}. In Section~\ref{sec:comparison}, we shall employ it to interpret the experiments described in Section~\ref{sec:materials}.

The elastic free energy density that was posited in \cite{paparini:elastic} has the distinguished feature of envisioning a \emph{double twist} (with two equivalent chiral variants) as ground state of CLCs in three-dimensional space. Here we briefly elaborate on this concept.

First, we introduce the independent scalar \emph{measures of distortion} $(S,T,B,q)$ for a director field $\n$; they are defined as follows:
$S:=\diver\n$ is the \emph{splay}, $T:=\n\cdot\curl\n$ is the \emph{twist}, $B:=|\bend|$ is the modulus of the \emph{bend} vector $\bend:=\n\times\curl\n$, and $q>0$ is the \emph{octupolar splay} \cite{pedrini:liquid}, derived from
\begin{equation}
	\label{eq:identity}
	2q^2=\tr(\nabla\n)^2+\frac12T^2-\frac12S^2.
\end{equation}  
Then, following a terminology proposed by Selinger~\cite{selinger:director} (see also \cite{long:explicit}), we call \emph{double} twist a distortion characterized by having \emph{all} measures of distortion equal to zero, \emph{but} a single one, $T$. Somehow ironically, a \emph{single} twist is characterized by having \emph{two} non-zero measures of distortion related to one another:
\begin{equation}
	\label{eq:single_twist}
	S=0,\quad B=0,\quad T=\pm2q.
\end{equation}
A single twist is nothing but what others would call a \emph{cholesteric} twist. While a single twist can fill \emph{uniformly} the whole space \cite{virga:uniform}, a double twist cannot. The latter can only be realized locally, not globally: it is thus a \emph{frustrated} ground state. It was shown in \cite{paparini:stability} that  a double twist can, for example, be attained exactly on the symmetry axis of cylinders enforcing degenerate planar anchoring on their lateral boundary.

The simplest way to induce locally a double twist is to add  a \emph{quartic twist} term to the classical Oseen-Frank  stored energy density \cite{oseen:theory,frank:theory} and define \cite{paparini:elastic}
\begin{equation}
	\label{eq:quartic_free_energy_density}
	\WQT(\n,\nabla\n):=\frac{1}{2}(K_{11}-K_{24})S^2+\frac{1}{2}(K_{22}-K_{24})T^2+ \frac{1}{2}K_{33}B^{2}+\frac{1}{2}K_{24}(2q)^2 + \frac14K_{22}a^2T^4,
\end{equation}
where $K_{11}$ is the \emph{splay} constant, $K_{22}$ is the \emph{twist} constant, $K_{33}$ is the \emph{bend} constant, $K_{24}$ the \emph{saddle-splay} constant, and $a$ is a \emph{characteristic length}. $\WQT$ is bounded below whenever 
\begin{subequations}\label{eq:new_inequalities}
	\begin{eqnarray}
		&K_{11}\geqq K_{24}\geqq0,\label{eq:new_inequalities_1}\\
		&K_{24}\geqq K_{22}\geqq0, \label{eq:new_inequalities_2}\\
		&K_{33}\geqq0,\label{eq:new_inequalities_3}
	\end{eqnarray}
\end{subequations}
which in this theory replace the celebrated \emph{Ericksen inequalities} \cite{ericksen:inequalities}. When, as we shall assume here, inequalities \eqref{eq:new_inequalities} hold strictly, $\WQT$ is minimum at the  \emph{degenerate} double twist characterized by
\begin{equation}
	\label{eq:double_twist}
	S = 0, \quad T = \pm T_0, \quad B = 0, \quad q = 0,
\end{equation}
where
\begin{equation}
	\label{eq:T_0min}
	T_0:=\frac{1}{a}\sqrt{\frac{K_{24}-K_{22}}{K_{22}}}.
\end{equation}
Here, we shall treat $a$ as a phenomenological parameter to be determined experimentally.
We designate the elastic theory based on $\WQT$ in \eqref{eq:quartic_free_energy_density} as quartic \emph{twist} theory, as among all possible quartic invariant terms in $\nabla\n$ we only consider the one proportional to $T^4$ (see also \cite{virga:uniform} for other possible quartic elastic theories).

\subsection{Twisted Hedgehog}\label{sec:hedgehog}
\begin{figure}
	\centering 
	\includegraphics[width=.33\linewidth]{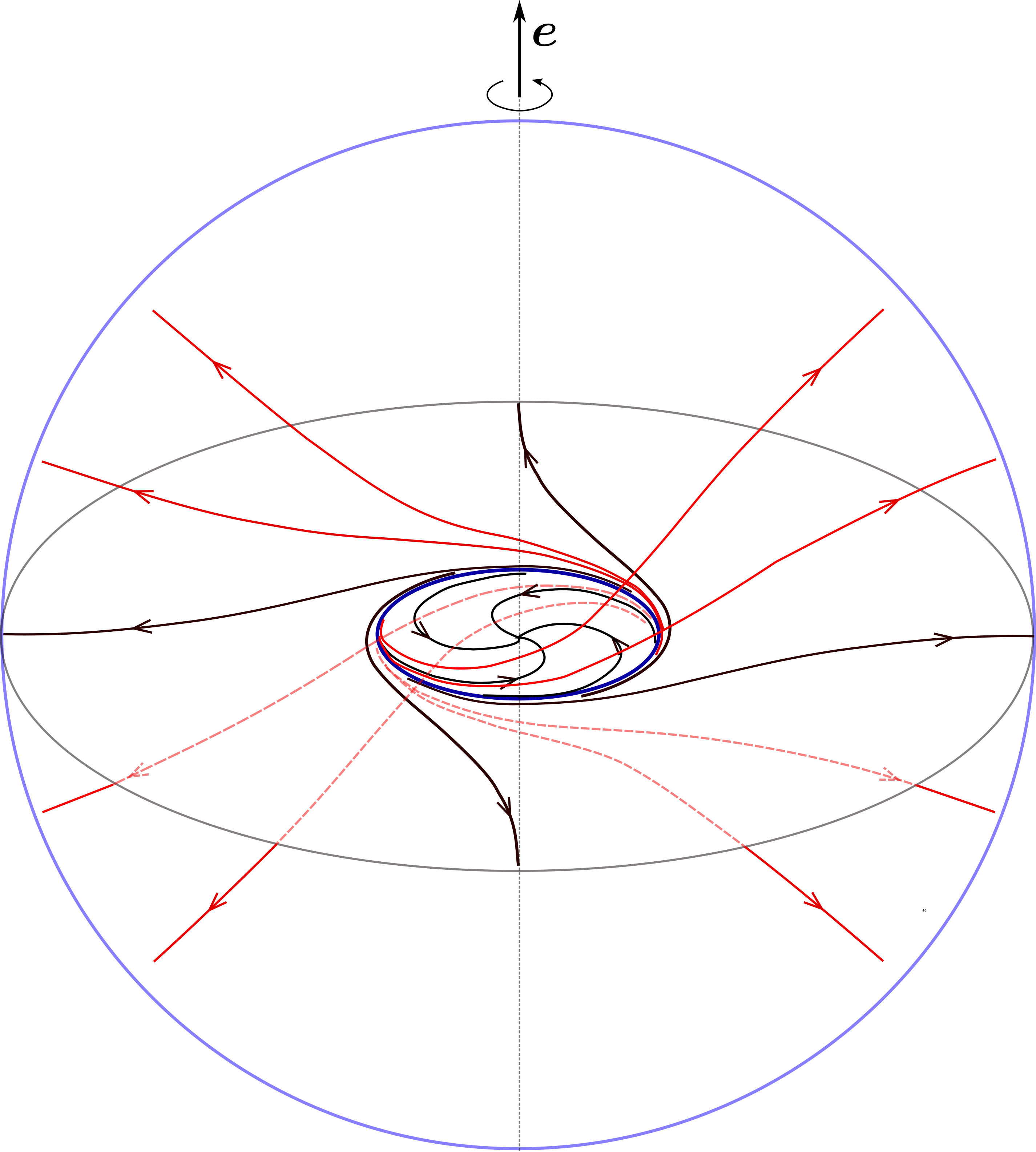}
	\caption{Field lines of $\nT$ in \eqref{eq:twisted_hedgehog} in a spherical cavity enforcing homeotropic boundary conditions. An inversion ring, depicted in blue, is present on the equatorial plane (orthogonal to the symmetry axis $\e$). Black lines are field lines lying on the equatorial plane; red lines are field lines coming out of the equatorial plane. The whole 3D picture is obtained by rotating this drawing about $\e$. This sketch  is adapted from Fig.~2 of \cite{paparini:spiralling}; it represent only one chiral variant of twisted hedgehog, the other is obtained by reversing the sign of $\alpha$ in \eqref{eq:twisted_hedgehog}.}
	\label{fig:twisted_hedgehog}
\end{figure}

The quartic twist theory was applied in \cite{paparini:spiralling} to describe the \emph{twisted hedgehog} that forms within a spherical cavity enforcing homeotropic alignment on its boundary. It is known since the seminal work of Lavrentovich and Terentiev \cite{lavrentovich:phase} that for a splay constant $K_{11}$ sufficiently larger than the twist constant $K_{22}$, the \emph{radial} hedgehog becomes unstable and acquires a twisted texture (with either sign of chirality equally likely to emerge), see Fig.~\ref{fig:twisted_hedgehog}. 
More precisely, as proved in \cite{cohen:weak,kinderlehrer:second,rudinger:twist}, the radial hedgehog loses local stability whenever
\begin{equation}
\label{eq:instability_inequality}
K_{11}>K_{22}+\frac18K_{33}.
\end{equation}
The original proof of this inequality was given for the Oseen-Frank elastic theory; however, it also applies to the theory associated with the energy density in  \eqref{eq:quartic_free_energy_density}, as both theories share the same second variation of the energy functional \cite{paparini:stability}. For CLCs, inequality \eqref{eq:instability_inequality} holds because $K_{11}$ and $K_{33}$ are customarily comparable, whereas $K_{22}$ is much smaller (nearly by an order of magnitude).

The field lines shown in Fig.~\ref{fig:twisted_hedgehog} were obtained in \cite{paparini:spiralling} by minimizing the total elastic free energy on the trial family of fields defined as
\begin{equation}
	\label{eq:twisted_hedgehog}
	\nT(\x):=\R(\alpha(r))\frac{\x}{r},
\end{equation}
where $\x$ is the position vector, $r:=|\x|$, and $\R(\alpha)$ denotes the rotation of angle $\alpha$ about a symmetry axis $\e$. Letting $\alpha$ depend only on $r$, we reduced the total elastic free energy to an admittedly complicated functional $\mathcal{F}[\alpha]$ of a single scalar function $\alpha(\rho)$, where $\rho:=r/R$ is the only space variable scaled to the radius $R$ of the spherical cavity. $\mathcal{F}$ is even, and so the same energy is assigned to a function $\alpha$ and its opposite. This degeneracy means that minimizers come in pairs of twisted hedgehogs with opposite chirality; Fig.~\ref{fig:twisted_hedgehog} represents only one chiral variant. Three dimensionless parameters feature in $\mathcal{F}$: the reduced elastic constants $k_1:=K_{11}/K_{22}$ and $k_3:=K_{33}/K_{22}$, and the ratio $\lambda:=a/R$. 

\begin{figure}
	\centering 
	\includegraphics[width=.4\linewidth]{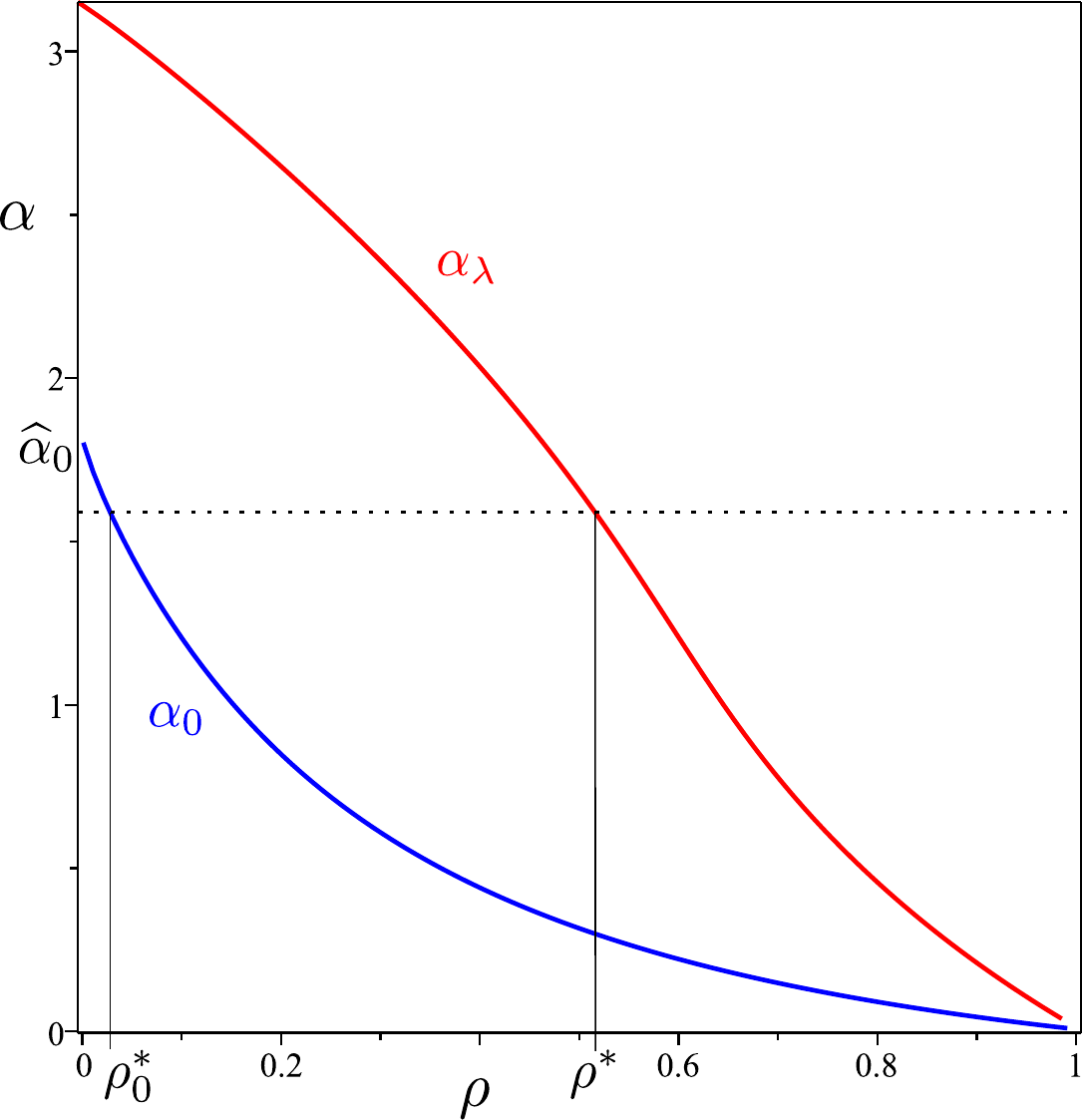}
	\caption{The elastic free energy (positive) minimizer $\alpha_\lambda$ for reduced elastic constants $k_1 = 7.5$,  $k_3 = 9.0$ (which apply to a SSY solution at concentration $c=30\,\mathrm{wt}\%$), and $\lambda=0.94$ is compared with the minimizer $\alpha_0$ according to the Oseen-Frank theory. The dotted line is drawn at $\alpha=\frac{\pi}{2}$; its intercepts with the graphs of $\alpha_\lambda$ and $\alpha_0$ designate the (scaled) radii of the inversion ring, $\rho^\ast:=r^\ast/R\doteq0.52$ and $\rho^\ast_0:=r^\ast_0/R\doteq0.03$, respectively. As shown in \cite{paparini:spiralling}, $\widehat{\alpha}_0=\arccos(-1/4)$ is the value of $\alpha_0$ at $\rho=0$, whereas $\alpha_\lambda(0)=\pi$.}
	\label{fig:alpha_profiles}
\end{figure}

In the special case where $\lambda=0$, the quartic twist theory reduces to the classical Oseen-Frank theory.  Figure~\ref{fig:alpha_profiles}
shows a typical positive minimizer $\alpha_\lambda$ of $\mathcal{F}[\alpha]$ compared with the minimizer $\alpha_0$ of the classical theory.

A remarkable feature of the twisted hedgehog is the \emph{inversion ring} that the nematic director field $\n$ exhibits on the plane orthogonal to the symmetry axis of the distortion texture. There, the direction of winding of the spiralling field lines of $\n$ changes; it has the optical appearance of a disclination  (see also Fig.~\ref{fig:cavities} below), but it is \emph{not} a defect, as  $\n$ is continuous there.

\begin{figure}
	\centering 
	\includegraphics[width=.4\linewidth]{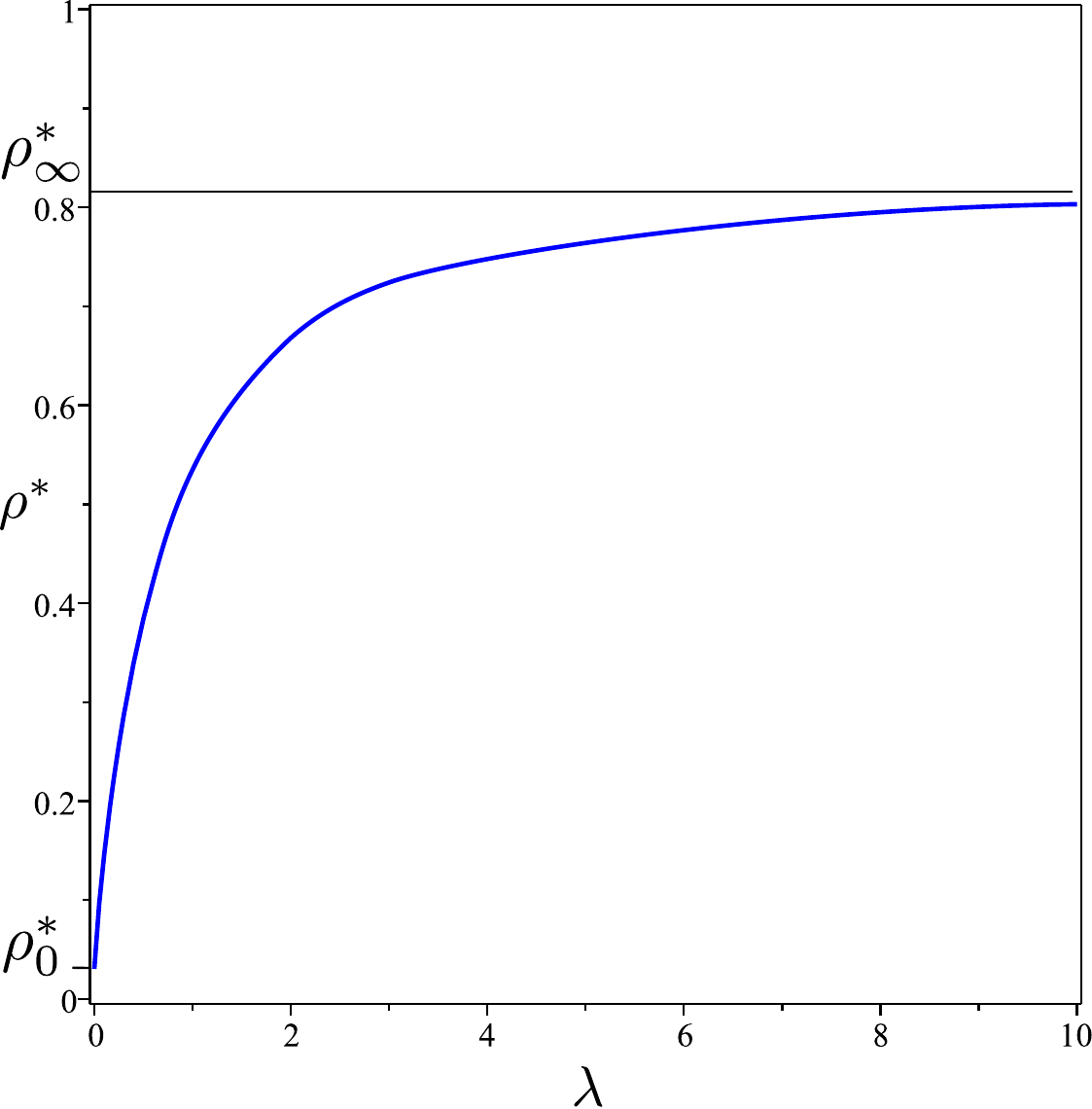}
	\caption{The scaled radius $\rho^\ast:=r^\ast/R$ of the inversion ring for the elastic free energy minimizer $\alpha_\lambda$ is plotted against $\lambda:=a/R$ for the same reduced elastic constants chosen in Fig.~\ref{fig:alpha_profiles}. The graph saturates at $\rho^*_\infty\doteq0.82$, while $\rho^*_0\doteq0.03$ is the limiting value as $\lambda\to0$.}
	\label{fig:inversion_ring_radius}
\end{figure}

How the radius $r^\ast$ of the inversion ring depends on $\lambda$ was studied in \cite{paparini:spiralling}. For given $\lambda$, $r^\ast$ can be read off from Fig.~\ref{fig:alpha_profiles}: it falls where the graph of $\alpha_\lambda$ intersects the line $\alpha=\frac{\pi}{2}$. The comparison between $\alpha_\lambda$ and  $\alpha_0$ suggests that the quartic twist theory may predict inversion rings appreciably larger than those compatible with the Oseen-Frank theory. Figure~\ref{fig:inversion_ring_radius} represents the graph of $\rho^\ast:=r^\ast/R$ against $\lambda$ for the same choice of reduced elastic constants made in Fig.~\ref{fig:alpha_profiles}.

It is the aim of this paper to provide an experimental validation of the theory recalled in this section. This will be achieved by extracting measures of the radius of the inversion ring produced in a  number of spherical cavities trapping a solution of Sunset yellow (SSY) in water inside a polymeric matrix enforcing homeotropic anchoring on the nematic director. How these cavities are obtained and with what materials is explained in the following section. 

\section{Experimental}\label{sec:materials}
\begin{figure}
	\centering
	\begin{subfigure}[c]{0.49\linewidth}
		\centering
		\includegraphics[width=.75\linewidth]{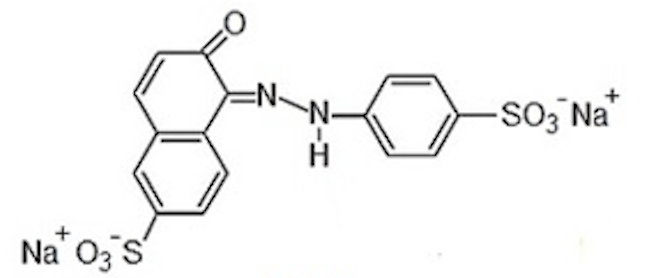}
		\caption{Molecular structure of SSY.}
\label{fig:1a}
	\end{subfigure}
	\begin{subfigure}[c]{0.49\linewidth}
		\centering
		\includegraphics[width=.9\linewidth]{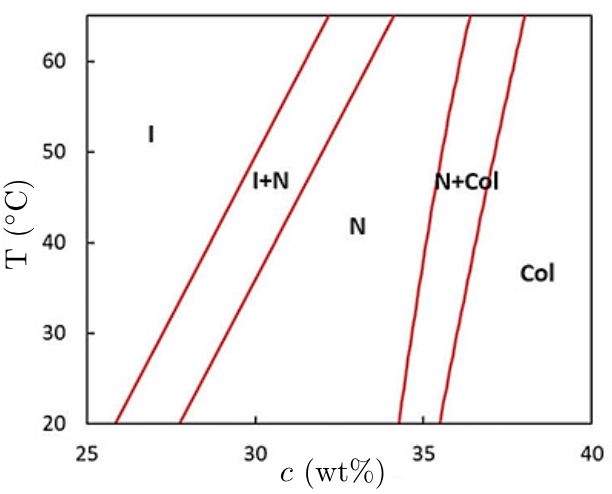}
	\caption{SSY phase diagram.}
\label{fig:1b}
	\end{subfigure}
\label{fig:1ab}
\caption{}
\end{figure}

The material SSY, whose molecular structure is shown in Fig.~\ref{fig:1a}, was purchased from Sigma-Aldrich and used without further purification. The brute formula is $\mathrm{C}_{16}\mathrm{H}_{10}\mathrm{N}_2\mathrm{Na}_{2}\mathrm{O}_7\mathrm{S}_2$ and the molecular weight is $452.37$. 

When dissolved in water it stacks in cylinder formations, the length of which depends on temperature and concentration; the cylinders then give rise to a nematic phase. At room temperature, the nematic phase exists between ca. $28$-$34\,\mathrm{wt}\%$ (Fig.~\ref{fig:1b}), \cite{bao:production}.

\begin{figure} 
\centering
	\includegraphics[width=0.25\linewidth]{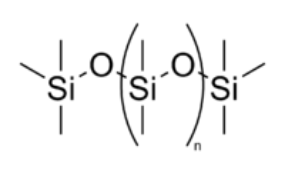}
\caption{Molecular structure of PDMS.}
\label{fig:2}
\end{figure}

SSY emulsions were prepared using polydimethylsiloxane (PDMS) as an immiscible matrix (Sylgard, kit $184$). PDMS is a liquid polymer that appears oily and transparent and has a high viscosity \cite{spina:intercalation}. To obtain a homeotropic anchoring at the interface, PDMS was used pure, as an oil in which to disperse the liquid crystal solution. PDMS brute formula is ($\mathrm{C}_2\mathrm{H}_6\mathrm{OSi})_n$, the structure of the repeating unit is shown in Fig.~\ref{fig:2}.

For the preparation of the chromonic solution, the SSY powder and water were mixed in appropriate proportions to obtain solutions containing SSY $30\, \mathrm{wt}\%$ and $31.5\,\mathrm{wt}\%$. 
The confinement of chromonic solutions in spherical geometries is achieved by preparing emulsions. The emulsions were prepared by adding a small amount of water-based chromonics solution to the oil matrix (PDMS).  By mechanically stirring, a large number of spherical microcavities were obtained. 

For microcavity texture investigations, a small amount of the prepared emulsion was sandwiched between two laboratory glass slides separated by mylar stripes of $190 \mu\mathrm{m}$. Due to the high viscosity of PDMS, it was not necessary to seal the cell with tape or glue.
Microsphere textures were studied using a polarized light microscope (Zeiss Axiolab $5$, Axiocam $208$ colour). Images were acquired with a monochromatic red filter whose transmission spectrum has peak at $639\,\mathrm{nm}$ (with full width at half maximum equal to $10\,\mathrm{nm}$). 

\section{Comparison with Theory}\label{sec:comparison}
\begin{figure} 
	\centering
	\includegraphics[width=0.5\linewidth]{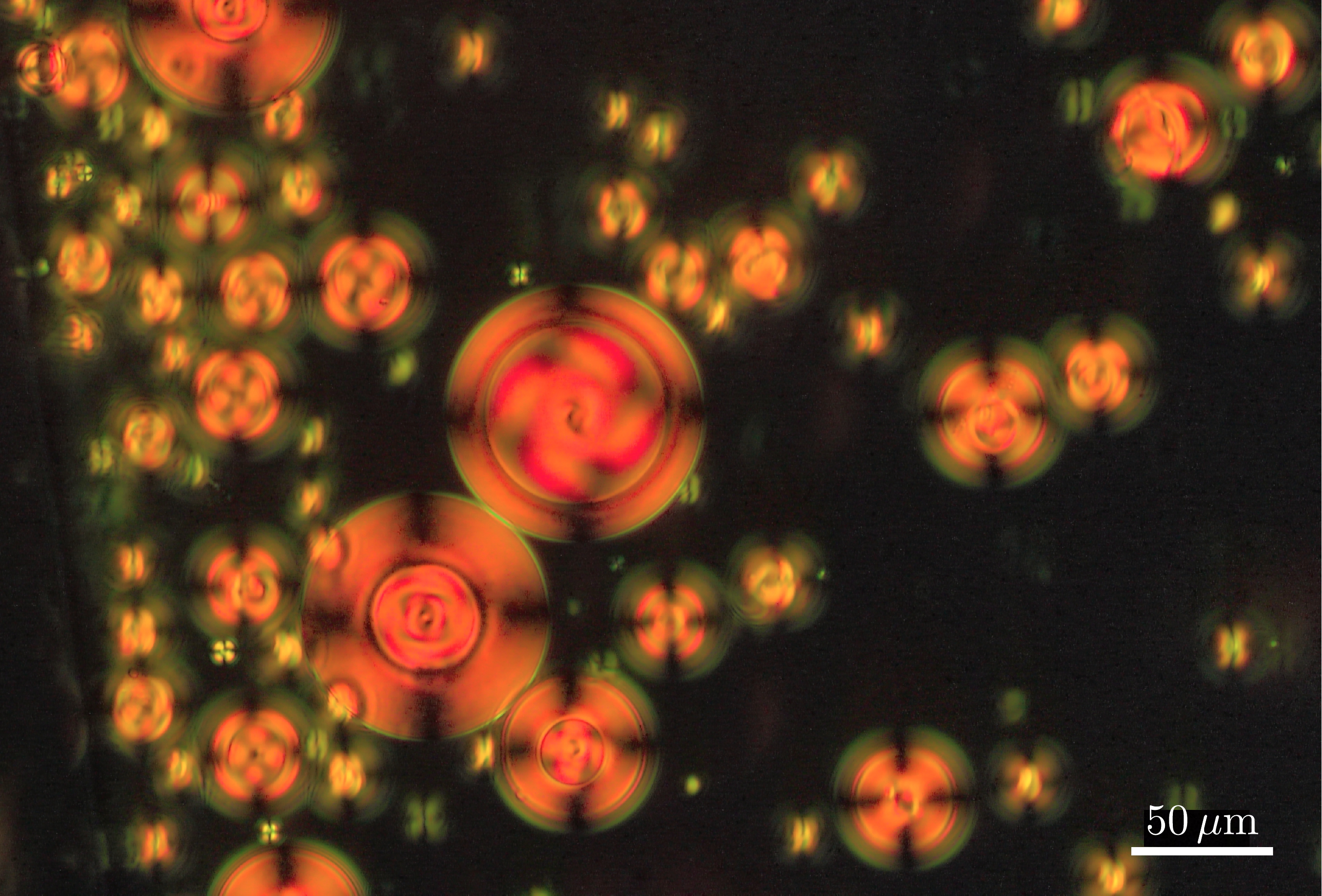}
	\caption{Microcavities enclosing a solution of SSY in water at concentration $c=30\,\mathrm{wt}\%$ and temperature  $\temp=25\,^\circ\mathrm{C}$. This particular sample was observed two days after preparation. All cavities allegedly host the same twisted hedgehog, but its symmetry axis is differently oriented  relative to the observer. For the cavities viewed along the hedgehog's symmetry axis, the inversion ring is easily identified as a circle; it does not look circular for the others, which are viewed askew  (see also Fig.~\ref{fig:confronto_img}).}
	\label{fig:cavities}
\end{figure}

A typical example of microcavities produced via the method described in Section~\ref{sec:materials} is shown in Figure~\ref{fig:cavities}.
These enclose a SSY solution in water at concentration $c=30\,\mathrm{wt}\%$ and temperature $\temp=25\,^\circ\mathrm{C}$. A similar experiment was also performed with a solution at $c=31.5\,\mathrm{wt}\%$ and same temperature. We conventionally refer to these as Experiment I and II, respectively. In both experiments, observations of the samples at the optical microscope were made immediately after having prepared the cell, as well as one and two days after. These are referred to as Observation $0$, $1$, and $2$, respectively; cavities seen in one observation are not the same as cavities seen in another.

\subsection{Inversion Ring Statistics}\label{sec:inversion_ring_statistics}
For each cavity that could be clearly discerned in images like the one in Fig.~\ref{fig:cavities}, we measured both $r^\ast$ and $R$ and then we extracted $a$ from the value of $\lambda=a/R$ read off from the graph like that in Fig.~\ref{fig:inversion_ring_radius} corresponding to the elastic constants of the material at the appropriate concentration and temperature. For the physical conditions of Experiment I, the reduced elastic constants are $k_1\approx7.5$ and $k_3\approx9.0$, while for those of Experiment II they are $k_1\approx11.0$ and $k_3\approx10.4$. (These values were derived from the measurements reported in  \cite{zhou:elasticity_2012,zhou:elasticity_2014,zhou:recent}, which we read as follows. Experiment I: $K_{11}\approx7.6\,\mathrm{pN}$, $K_{22}\approx1.02\,\mathrm{pN}$, $K_{33}\approx9.2\,\mathrm{pN}$; Experiment II: $K_{11}\approx13.8\,\mathrm{pN}$, $K_{22}\approx1.25\,\mathrm{pN}$, $K_{33}\approx13.0\,\mathrm{pN}$.)

\begin{figure}
	\centering
	\begin{subfigure}[c]{0.37\linewidth}
		\centering
		\includegraphics[width=\linewidth]{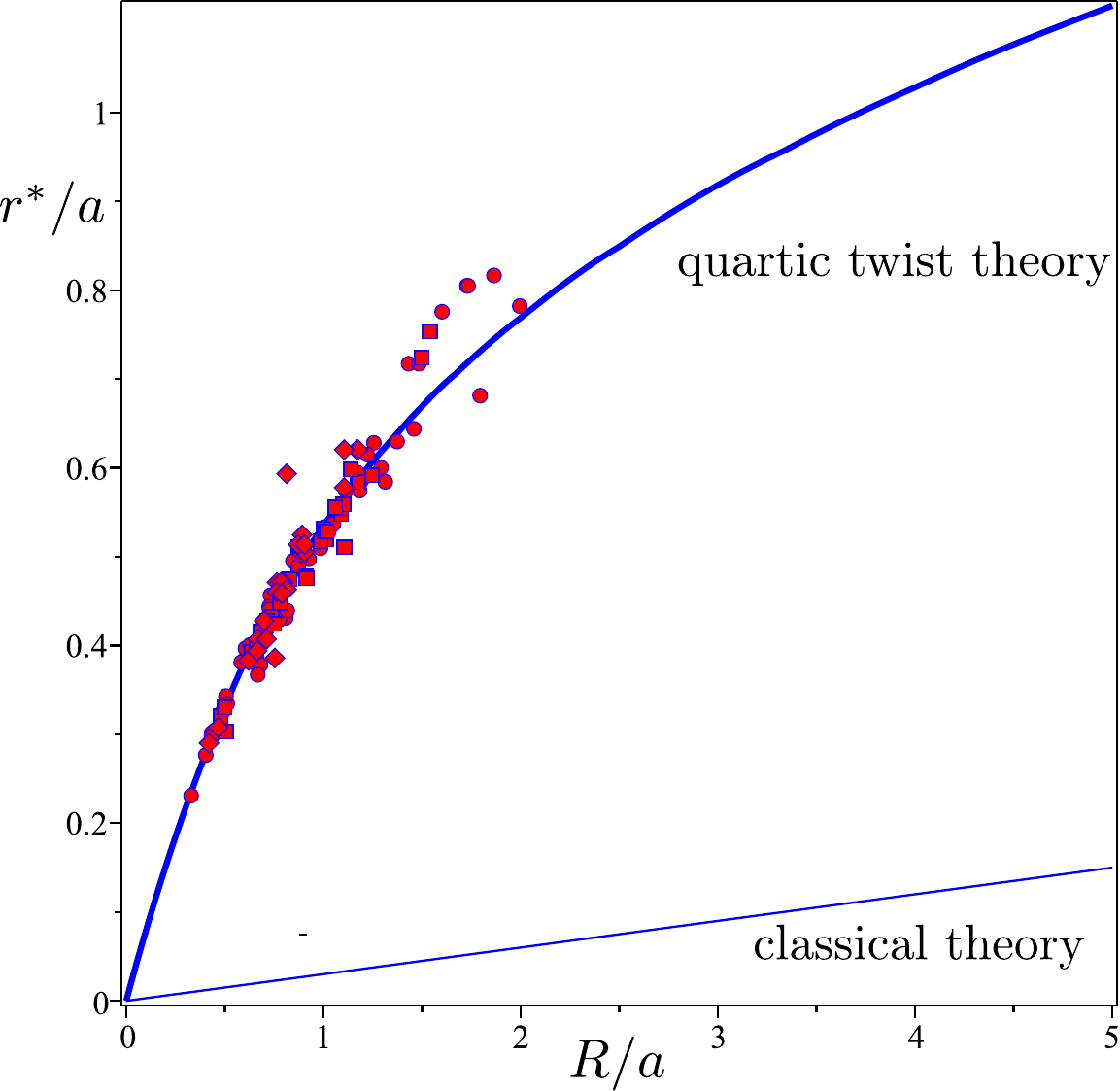}
		\caption{Experiment I: SSY solution at concentration $c=30\,\mathrm{wt}\%$  and temperature $\temp=25\,^\circ\mathrm{C}$.
			The total number of data points is $116$.}
		\label{fig:r_star_vs_R_30}
	\end{subfigure}
	$\qquad\ $
	\begin{subfigure}[t]{0.15\linewidth}
		\centering
		\includegraphics[width=\linewidth]{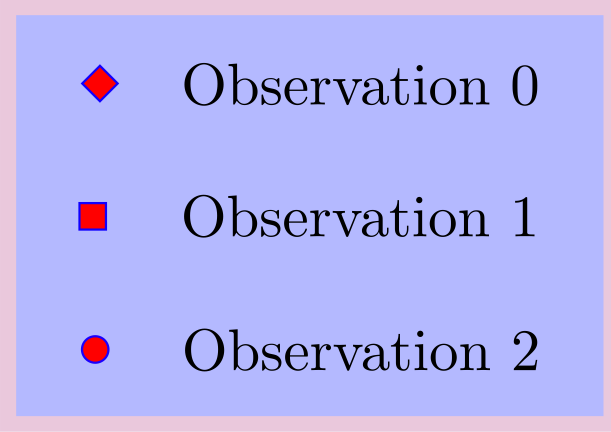}
	\end{subfigure}
	\begin{subfigure}[c]{0.40\linewidth}
		\centering
		\includegraphics[width=0.95\linewidth]{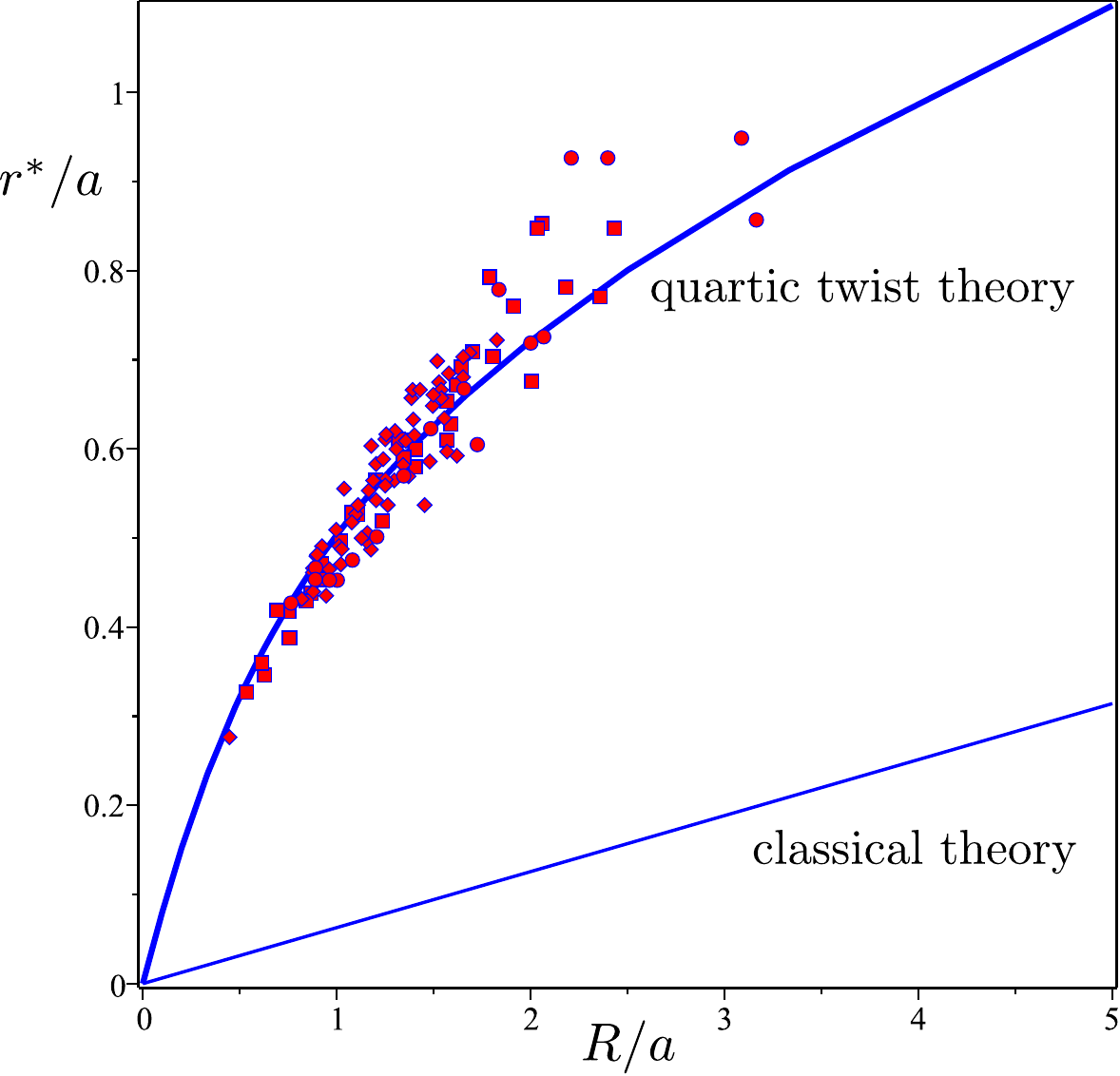}
		\caption{Experiment II: SSY solution at concentration $c=31.5\,\mathrm{wt}\%$ and temperature $\temp=25\,^\circ\mathrm{C}$.
			The total number of data points is $114$.}
		\label{fig:r_star_vs_R_31_5}
	\end{subfigure}
	\caption{Data collected in all observations. They are shown against the universal graph of the function $r^\ast/a$ against $R/a=1/\lambda$ obtained from the function $\rho^\ast$ plotted in Fig.~\ref{fig:inversion_ring_radius} against $\lambda$ via the transformation \eqref{eq:axes_transformation}. Diamonds refer to samples just prepared (Observation $0$), boxes to  samples observed after one day (Observation $1$),   dots to  samples observed after two days (Observation $2$). Straight lines apply to the classical Oseen-Frank theory; they are recovered asymptotically by the quartic twist theory in the limit as $R/a\to\infty$.}
	\label{fig:r_star_vs_R}
\end{figure}

The data thus collected for $r^\ast$, $R$, and $a$ were more conveniently organized on a universal graph (depending only on $k_1$ and $k_3$) that plots the ratio $r^\ast/a$ against $R/a=1/\lambda$. This graph is different for each experiment; it is  obtained from the graph in Fig.~\ref{fig:inversion_ring_radius} via the transformation
\begin{equation}
	\label{eq:axes_transformation}
	\frac{r^\ast}{a}=\frac{\rho^\ast}{\lambda}.
\end{equation}
This changes the function $\rho^\ast$ plotted in Fig.~\ref{fig:inversion_ring_radius} against $\lambda$ into a monotonic function that approaches $0$ as $R/a\to0$ and grows nonlineraly until becoming asymptotically linear for large $R/a$, $r^\ast/R\approx\rho^\ast_0(R/a)$,  where $\rho^\ast_0$ is the (scaled) radius of the inversion ring predicted by the Oseen-Frank theory. The rearranged data are shown in Fig.~\ref{fig:r_star_vs_R} for both experiments.
As also recalled in \cite{paparini:spiralling}, the classical Oseen-Frank theory predicts an inversion ring whose radius $r^\ast_0$ is such that the ratio $r^\ast_0/R$ depends only on the reduced elastic constants; thus, the graphs representing the prediction of the classical theory in both panels of Fig.~\ref{fig:r_star_vs_R} are straight lines. More precisely, for Experiment I,  $r^*_0/R=\rho^*_0\approx0.03$, while $r^*_0/R=\rho^*_0\approx0.06$ for Experiment II. It is apparent that for chromonics the Oseen-Frank theory fails to represent our data, whereas the proposed quartic twist theory seems in good agreement with both experiments, more so for smaller cavities than for larger ones, for which the inversion ring tends to be larger than predicted by theory.

All collected data are shown in Fig.~\ref{fig:r_star_vs_R}: they include all observations.
The reader will find in Appendix~\ref{sec:supplementary_data} separate illustrations for each observation in both experiments. 

\begin{figure} 
	\centering
	\includegraphics[width=0.45\linewidth]{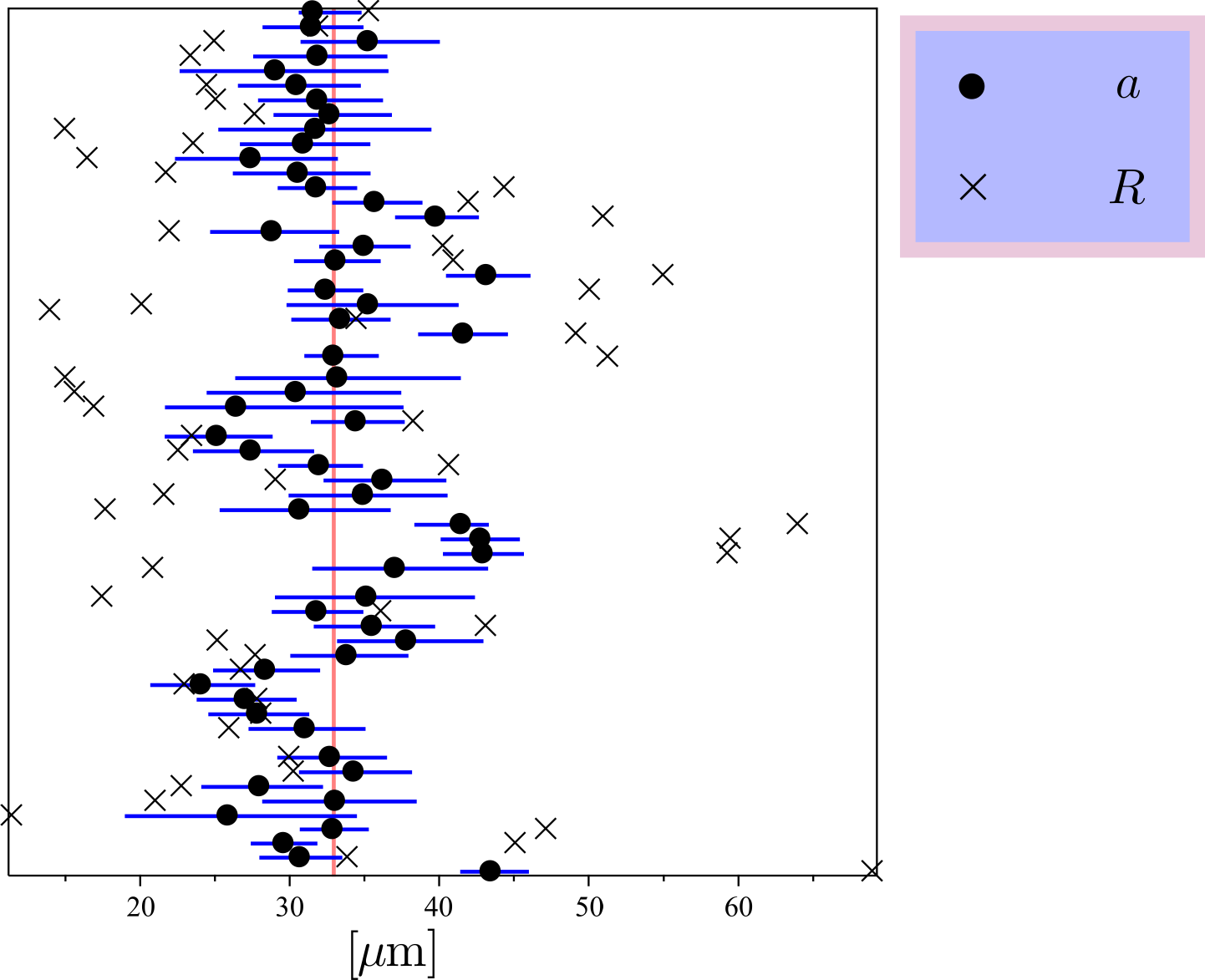}
	\caption{Dispersion histogram of the data in Fig.~\ref{fig:r_star_vs_R_30} for the observation two days after preparation in Experiment I. The measured values $R_i$ of the radius $R$  are black crosses and the estimated values $a_i$ of the characteristic length $a$ are black dots  (with corresponding blue error bars). The red vertical line marks the mean value $\bar{a}\approx33\,\mu\mathrm{m}$ of $a$.}
	\label{fig:dispersion_histogram}
\end{figure}

Since the measures of both $r^\ast$ and $R$ are affected by errors, these propagate to $a$, in a way made possible to estimate by the  predicted dependence of $\rho^\ast$ on $\lambda$ (shown in Fig.~\ref{fig:inversion_ring_radius}). For each observation, we also estimated how the inferred values of $a$ are distributed compared to the measured values of the radius $R$. Our goal is to show that the former are less dispersed than the latter, as appropriate for a phenomenological length that should only be related to the material and its physical conditions. The comparison is made easier by the \emph{dispersion} histogram shown in Fig.~\ref{fig:dispersion_histogram} for a single observation (at a given SSY concentration).
For completeness, similar dispersion histograms for all other observations are collected in Appendix~\ref{sec:error_analysis}. Despite the noise present in the data, we can be reasonably confident in the material nature of $a$. Quantitative details on how we determined the error bars in Fig.~\ref{fig:dispersion_histogram} are also given in Appendix~\ref{sec:error_analysis}.

\subsection{Estimates of $a$}\label{sec:estimate_a}
The method illustrated in Section~\ref{sec:inversion_ring_statistics} led us to the estimates of $a$ collected in Table~\ref{tab:a}.
\begin{table}
\begin{center}
\begin{tabular}{|c|c|c|c| }
\cellcolor{lightgray}{SSY concentration} & \cellcolor{lightgray}{Observation $0$}  & \cellcolor{lightgray}{Observation $1$} & \cellcolor{lightgray}{Observation $2$}  \\ 
\rule{0pt}{2ex}
$30\,\mathrm{wt}\%$  &  $a\approx 47 \, \mu\mathrm{m} $ &  $a\approx 41 \, \mu\mathrm{m} $ &  $a\approx 33 \, \mu\mathrm{m} $ \\
\hline
$31.5\,\mathrm{wt}\%$ & $a\approx 54 \, \mu\mathrm{m} $ &  $a\approx 46 \, \mu\mathrm{m} $ &  $a\approx 35 \, \mu\mathrm{m}$ \\
 \hline
\end{tabular}
\end{center}
\caption{Estimates of $a$ for SSY solutions in the physical conditions of different  observations.}
\label{tab:a}
\end{table}
The corresponding deviations from the mean values (which do not exceed $9\%$) are  recorded in Appendix~\ref{sec:error_analysis}, which also documents the much larger dispersion in the data set for $R$.

It is worth remarking that for SSY in the same physical conditions as in Experiment I, in \cite{paparini:elastic} we had estimated $a\approx6.4\,\mu\mathrm{m}$ from the director field observations of \cite{davidson:chiral} in cylinders with planar degenerate anchoring conditions on their lateral boundary. The difference (by nearly one order of magnitude) between this estimate and those summarized in Table~\ref{tab:a} might result from the different experimental settings. By the disparity in the number of data collected here and in \cite{davidson:chiral} and the fact that in \cite{paparini:elastic} we needed to determine two fitting parameters ($a$ and $K_{24}$) instead of one, we are inclined to think that the estimates summarized in Table~\ref{tab:a} are more reliable.
 
\begin{figure} 
\centering
	\includegraphics[width=0.4\linewidth]{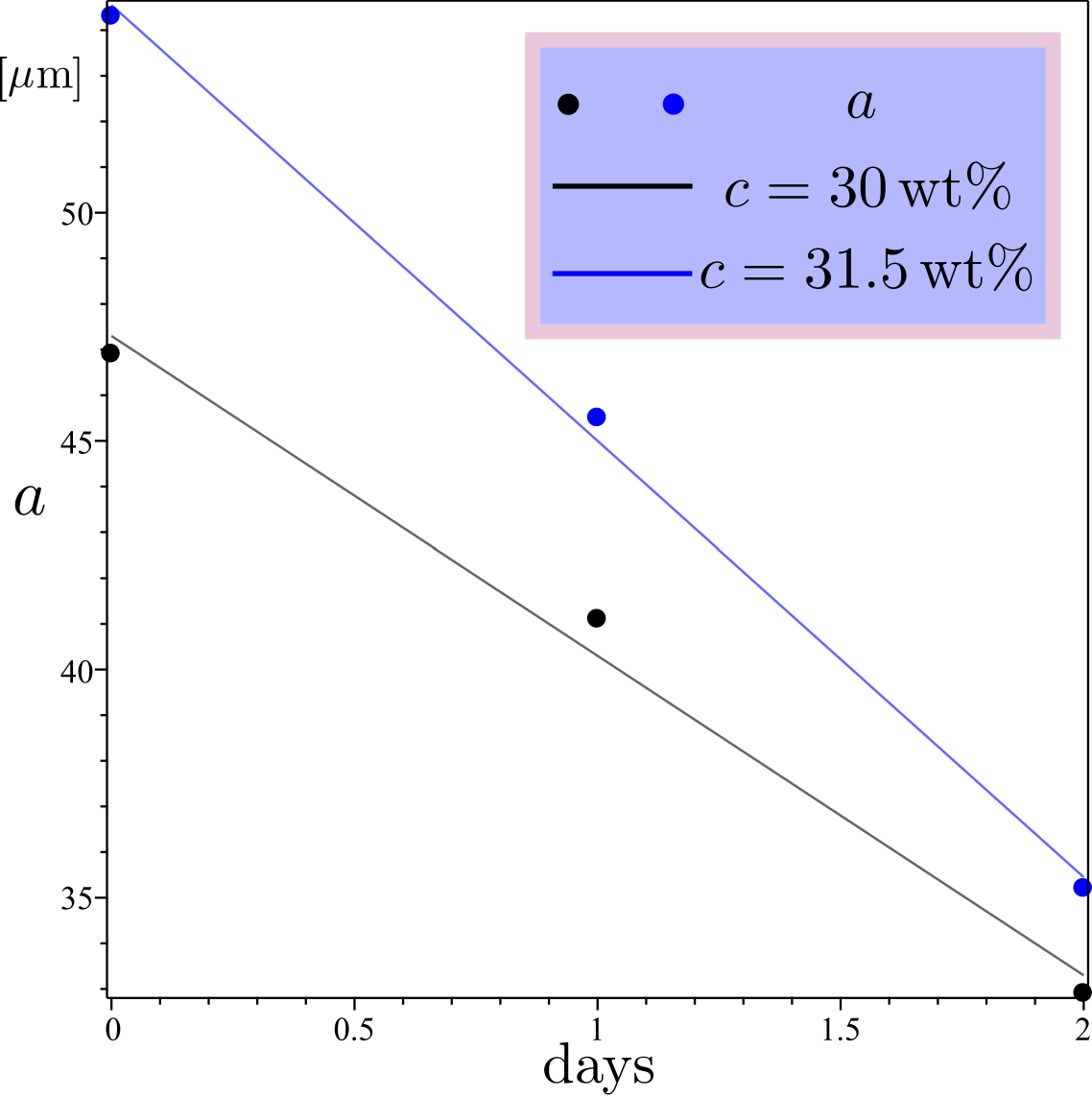}
\caption{Estimated values  of $a$ plotted against the time elapsed since cell preparation for both SSY concentrations, namely, $c=30\,\mathrm{wt}\%$ (black line) and $c=31.5\,\mathrm{wt}\%$ (blue line). Dots mark the values in Table \ref{tab:a}; straight lines are guides for the eye.}
\label{fig:a_bar_after_two_days}
\end{figure}

Two trends emerge clearly from these findings: $a$ increases with concentration and decreases as time  elapses since cell preparation (see Fig.~\ref{fig:a_bar_after_two_days}).

\subsection{Spiralling Cores}\label{sec:cores} 
As proved in \cite{paparini:spiralling} and illustrated in Fig.~\ref{fig:twisted_hedgehog}, the quartic twist theory recalled in Section~\ref{sec:theory} predicts a distortion \emph{spiralling core} within the inversion ring of a twisted hedgehog. Here we compare with experiment this qualitative feature of the theory. 

\begin{figure}
	\centering
	\begin{subfigure}[c]{0.45\linewidth}
		\centering
		\includegraphics[width=\linewidth]{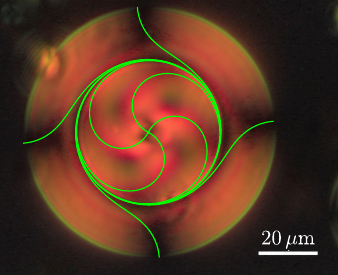}
		\caption{Spherical cavity of radius $R\approx 44\,\mu\mathrm{m}$ observed after one day from the sample preparation. Taking $a\approx41\,\mu\mathrm{m}$ from Table \ref{tab:a}, we estimate $r^*\approx23\,\mu\mathrm{m}$.}
\label{fig:campione2_9_1gg_field}
	\end{subfigure}
	$\quad$
	\begin{subfigure}[c]{0.45\linewidth}
		\centering
		\includegraphics[width=.88\linewidth]{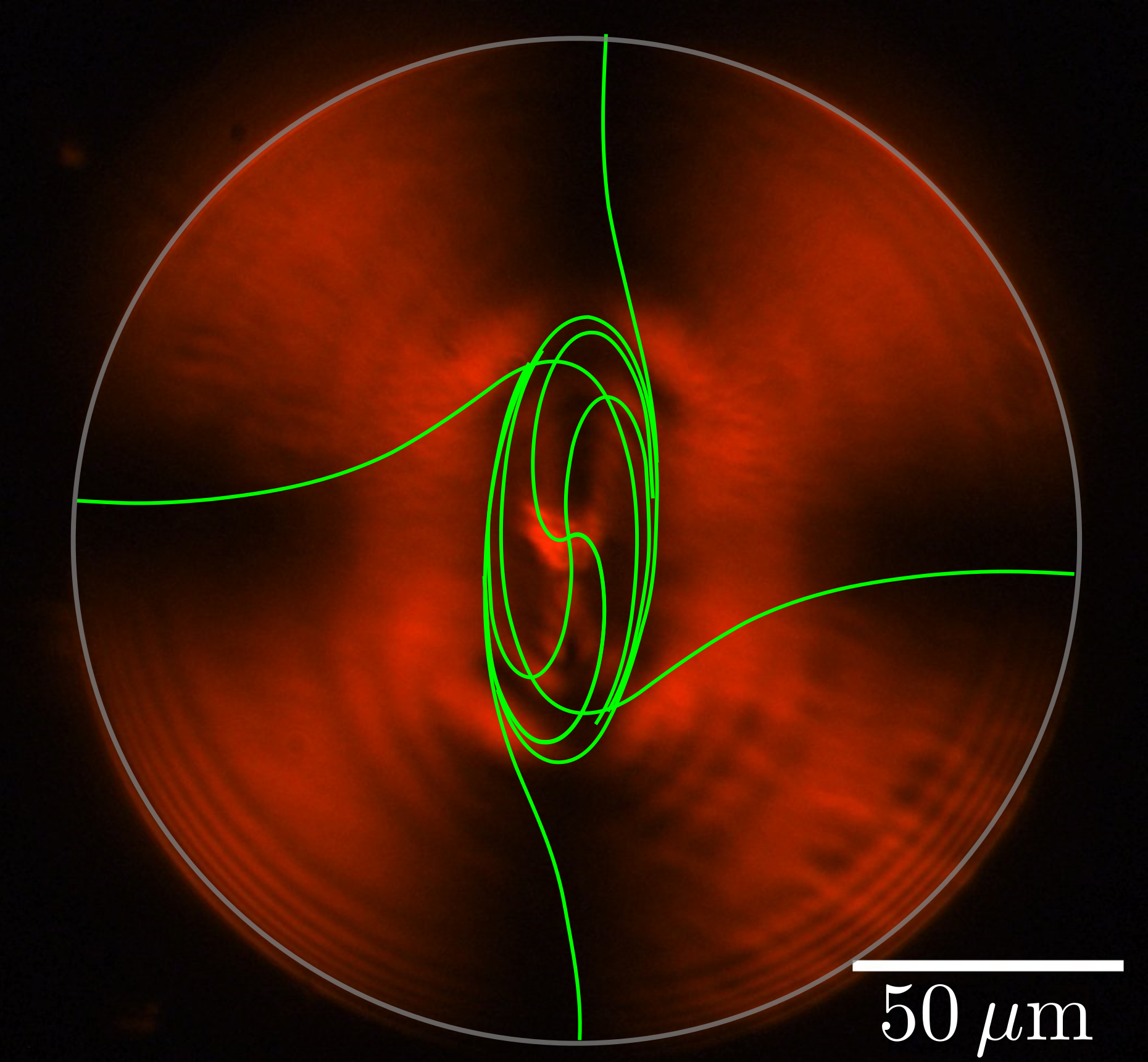}
	\caption{Spherical cavity of radius $R\approx 105\mu\mathrm{m}$ observed in a just prepared sample. Taking $a\approx47\,\mu\mathrm{m}$ from Table \ref{tab:a}, we estimate $r^*\approx46\,\mu\mathrm{m}$.}
\label{fig:confronto_img}
	\end{subfigure}
\caption{Spherical cavities enclosing a SSY solution at concentration $c=30\,\mathrm{wt}\%$ and temperature $\temp=25\,^\circ\mathrm{C}$. The spiralling core predicted by  the  quartic twist theory (depicted in green) is superimposed to the experimental images. The radius of the inversion ring is estimated by theory (as explained in Section~\ref{sec:theory}); its predicted values are in good agreement with direct measures.}
\label{fig:droplets_experiments}
\end{figure}

Among the many observed cavities, we focus on two exemplary cases; they are shown in Fig.~\ref{fig:droplets_experiments}. 
In both cases, SSY is at concentration $c=30\,\mathrm{wt}\%$ and temperature $\temp=25\,^\circ\mathrm{C}$: the cavity in Fig.~\ref{fig:campione2_9_1gg_field}
was observed  one day after preparation, while the cavity in Fig.~\ref{fig:confronto_img} was observed just at the time of preparation. In the former instance, the spiralling core is observed along the symmetry axis of the twisted hedgehog (at right angles with the plane containing the inversion ring), whereas in the latter it is viewed askew, along a direction that we estimated at the angle $\pi/8$ from the symmetry axis.

Based on this evidence, we may say that the quartic twist theory seems also to be in good qualitative agreement with the experimental observations.

\section{Conclusions}\label{sec:conclusion}
A quartic twist theory for the elasticity of chromomic liquid crystals is subjected to experimental scrutiny. The spiralling texture of twisted hedgehogs in spherical cavities enforcing homeotropic anchoring is characterized by an inversion ring, which can be observed optically. Measurements of the inversion ring are contrasted with both the classical Oseen-Frank theory and the quartic twist theory, and shown to be in better accord with the latter than with the former.

A phenomenological length $a$ features in the proposed theory; it is a material parameter, supposedly depending on both temperature and concentration of the chromonic solution. The data collected for the inversion ring agree with theory if a specific value of $a$ is chosen for each cavity's radius $R$. The small dispersion in the values of $a$ compared to the dispersion of radii is a testament to the material nature of $a$, which depends only on the physical conditions in which the observation takes place.

The average of $a$ for each observation is found to increase with concentration and decrease with the time elapsed after sample preparation. The former trend can be understood by recalling that a similar behaviour is shown by the ratio of elastic constants $K_{11}/K_{33}$ in homogeneous chromonic phases (see either Fig.~3 of \cite{zhou:elasticity_2012} or Fig.~2c of \cite{zhou:recent}). According to the microscopic theory developed in \cite{taratuta:anisotropic} (see also \cite{odijk:elastic}) for the elastic properties of liquid crystal phases generated by \emph{flexible} molecules,
\begin{equation}
	\label{eq:constant_ratio}
	\frac{K_{11}}{K_{33}}=\frac{\ell}{\delta},
\end{equation}  
where $\ell$ is the \emph{contour} length of the constituting molecules and $\delta$ is their \emph{persistence} length. In the present context, since $\delta$ is far less sensitive to concentration than $\ell$, \eqref{eq:constant_ratio} would suggest that $\ell$ (here the length of CLC aggregates) increases with $c$. Since $a$ increases with $c$ too, one is led to conjecture that $a$ is an increasing function of $\ell$.

This is, however, a slippery terrain, as observations of twisted tactoids in the biphasic regime suggested in \cite{nayani:using} that $\ell$ might instead be a decreasing function of $c$, at least for temperatures just slightly below the biphasic-to-isotropic transition.\footnote{The droplets observed in \cite{nayani:using} are in the biphasic region of phase space, where nematic and isotropic phases of the solution coexist in equilibrium, although at different concentrations, higher in the nematic ($c_\mathrm{N}$), lower in the isotropic ($c_\mathrm{I}$); see Fig.~\ref{fig:1b}. If  in \cite{nayani:using} $c>c_\mathrm{I}$ is the concentration at which the solution is prepared, then the concentration in the twisted tactoids would  be $c_\mathrm{N}>c$.}

In our experiments, temperature was not varied, but (different) cavities were observed at different times. The length $a$ was found to decrease as such a \emph{curing} time elapsed. If the conjectured link between $a$ and $\ell$ is to be trusted, one may interpret this result as suggesting that molecular aggregates become shorter on average (hence larger in number) as the cure progresses. A theoretical validation of this conjecture could be found by building a kinetic model for chromonic molecular aggregation, possibly along lines similar to  those that started to be traced in \cite{pergamenshchik:kinetic,pergamenshchik:statistical}.

\section*{Acknowledgments}
The UNICAL Group acknowledges financial support through the Italian PRIN 2022 - PNRR Project No. P2022HM5E4 on \emph{Chirality induction in water based self-assembling materials}. 

\appendix

\section{Data Disaggregation}\label{sec:supplementary_data}
\begin{figure}
	\centering
	\begin{subfigure}[c]{0.30\linewidth}
		\centering
		\includegraphics[width=\linewidth]{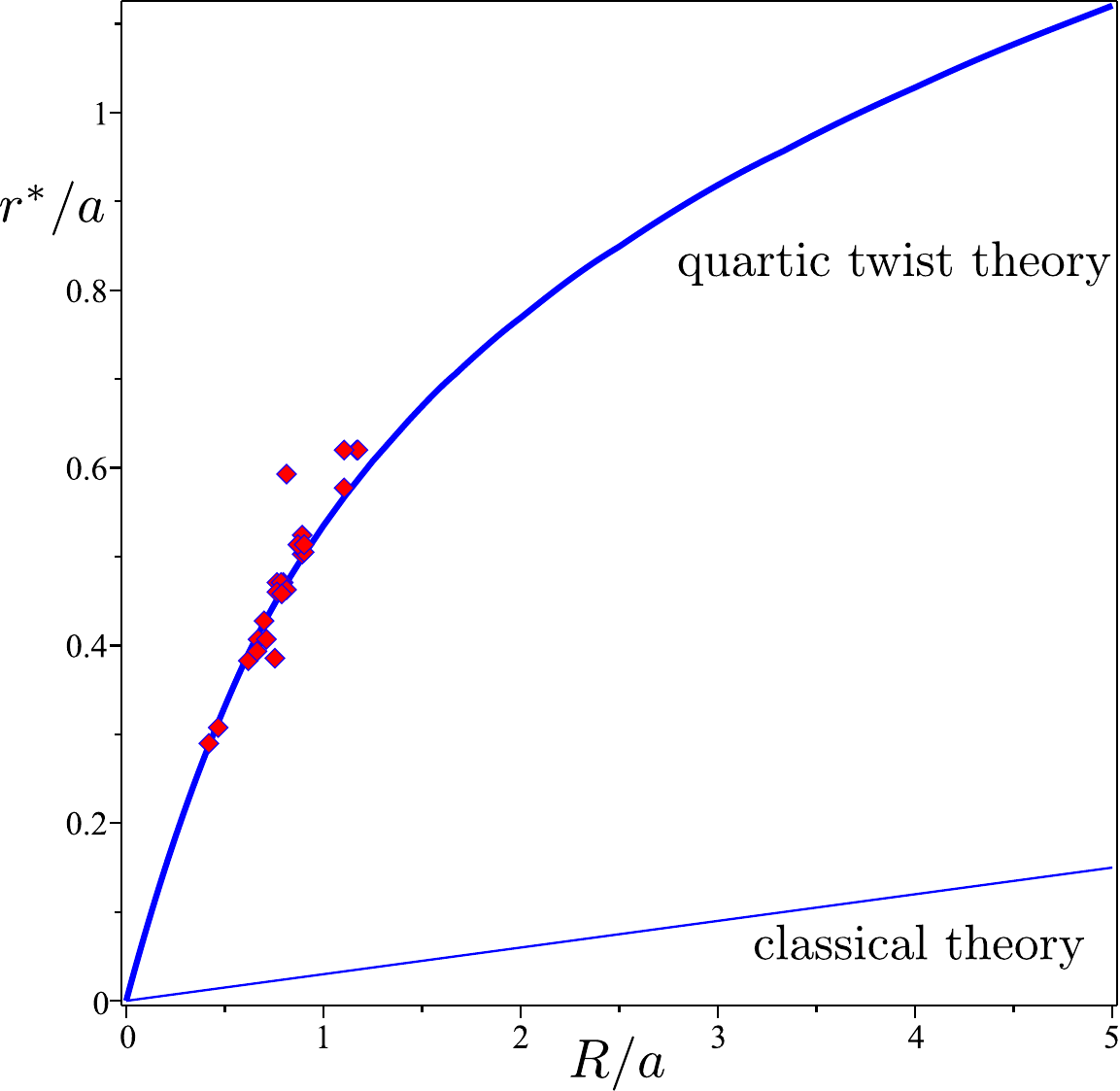}
		\caption{Experiment I; Observation 0.} 
		\label{fig:universal_I_0}
	\end{subfigure}
	\begin{subfigure}[c]{0.3\linewidth}
		\centering
		\includegraphics[width=\linewidth]{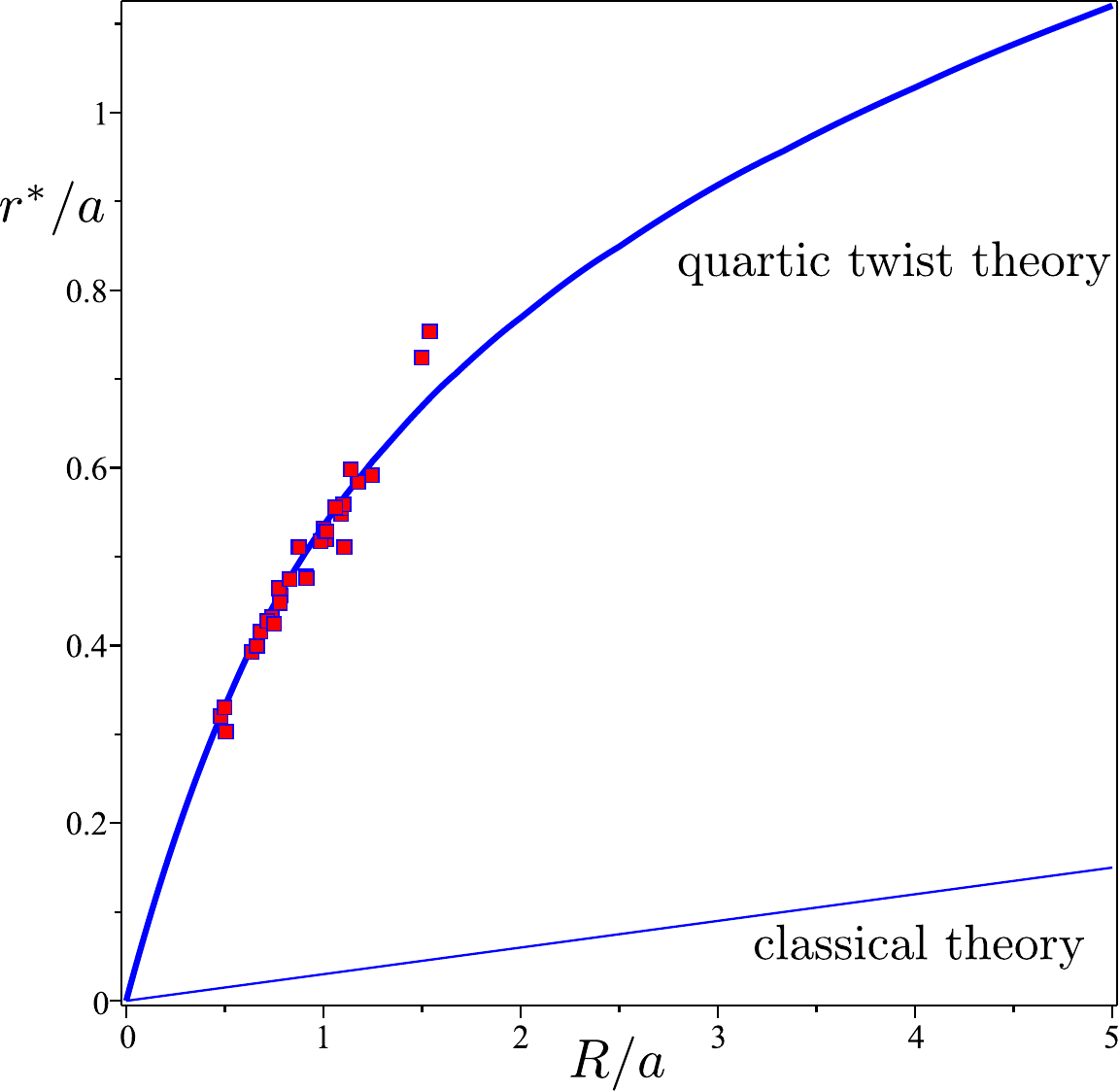}
		\caption{Experiment I; Observation 1.}
		\label{fig:universal_I_1}
	\end{subfigure}
	\begin{subfigure}[c]{0.3\linewidth}
		\centering
		\includegraphics[width=\linewidth]{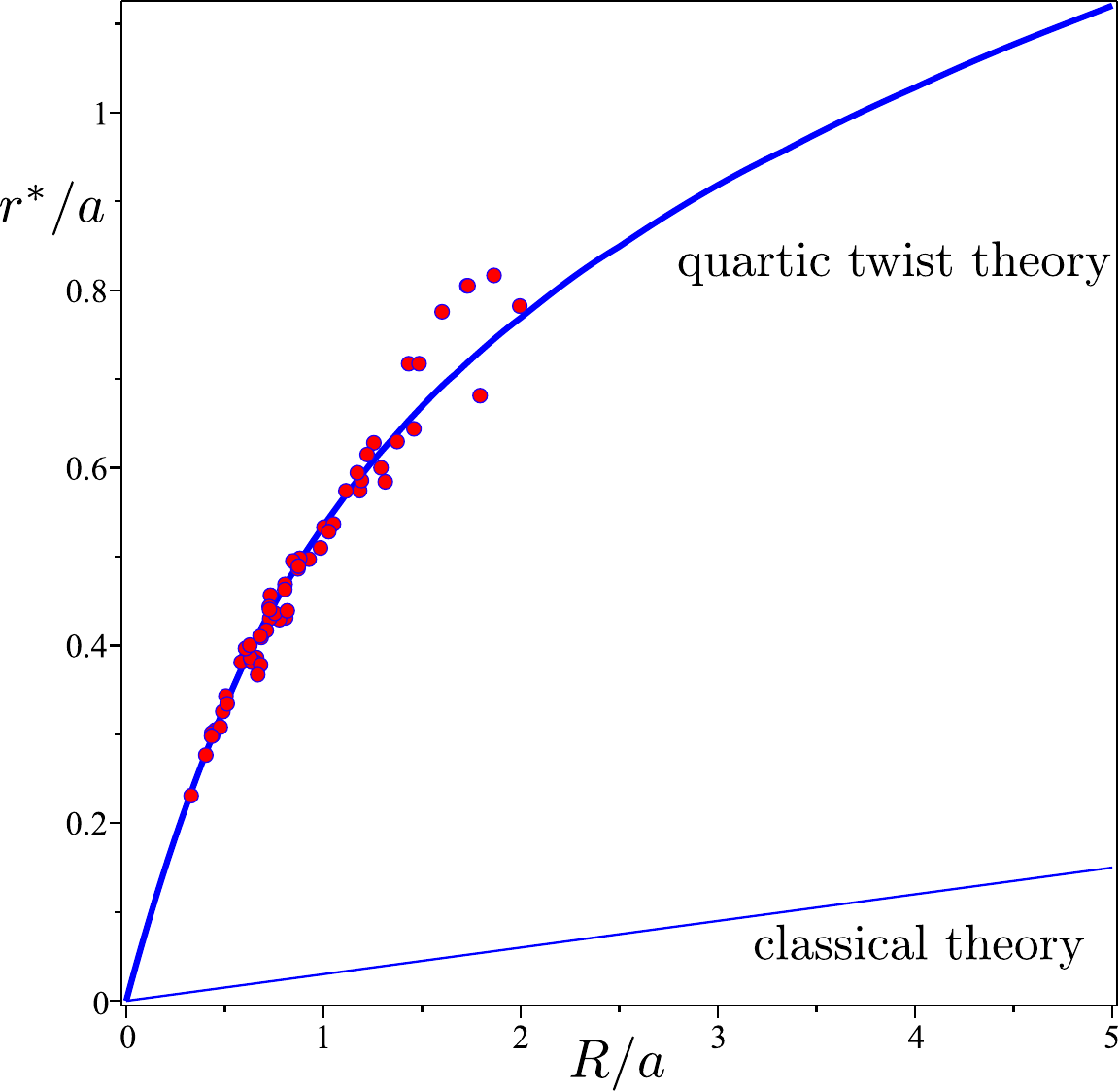}
		\caption{Experiment I; Observation 2.}
		\label{fig:universal_I_2}
	\end{subfigure}
	\begin{subfigure}[c]{0.30\linewidth}
		\centering
		\includegraphics[width=\linewidth]{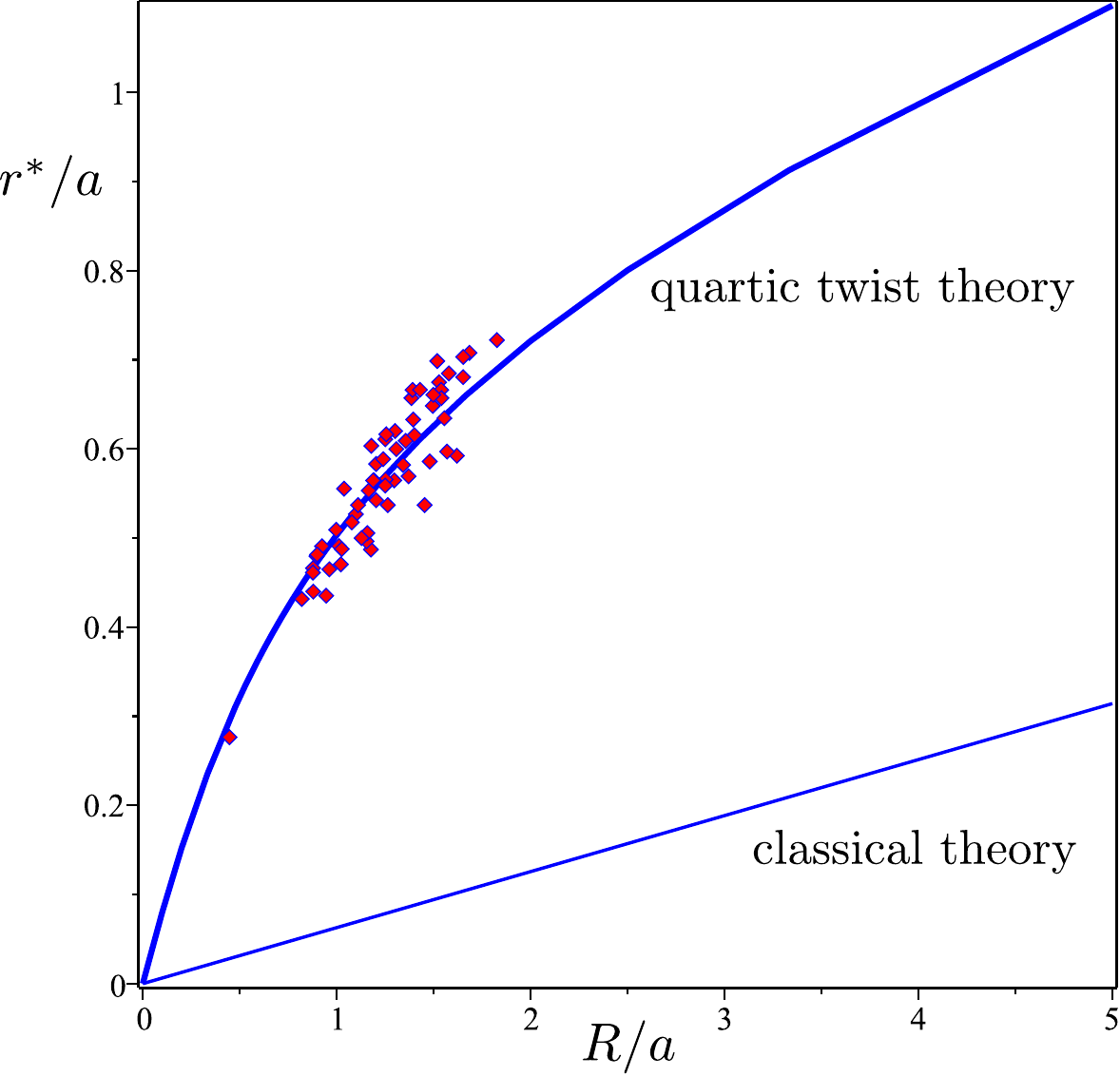}
		\caption{Experiment II; Observation 0.} 
		\label{fig:universal_II_0}
	\end{subfigure}
	\begin{subfigure}[c]{0.30\linewidth}
		\centering
		\includegraphics[width=\linewidth]{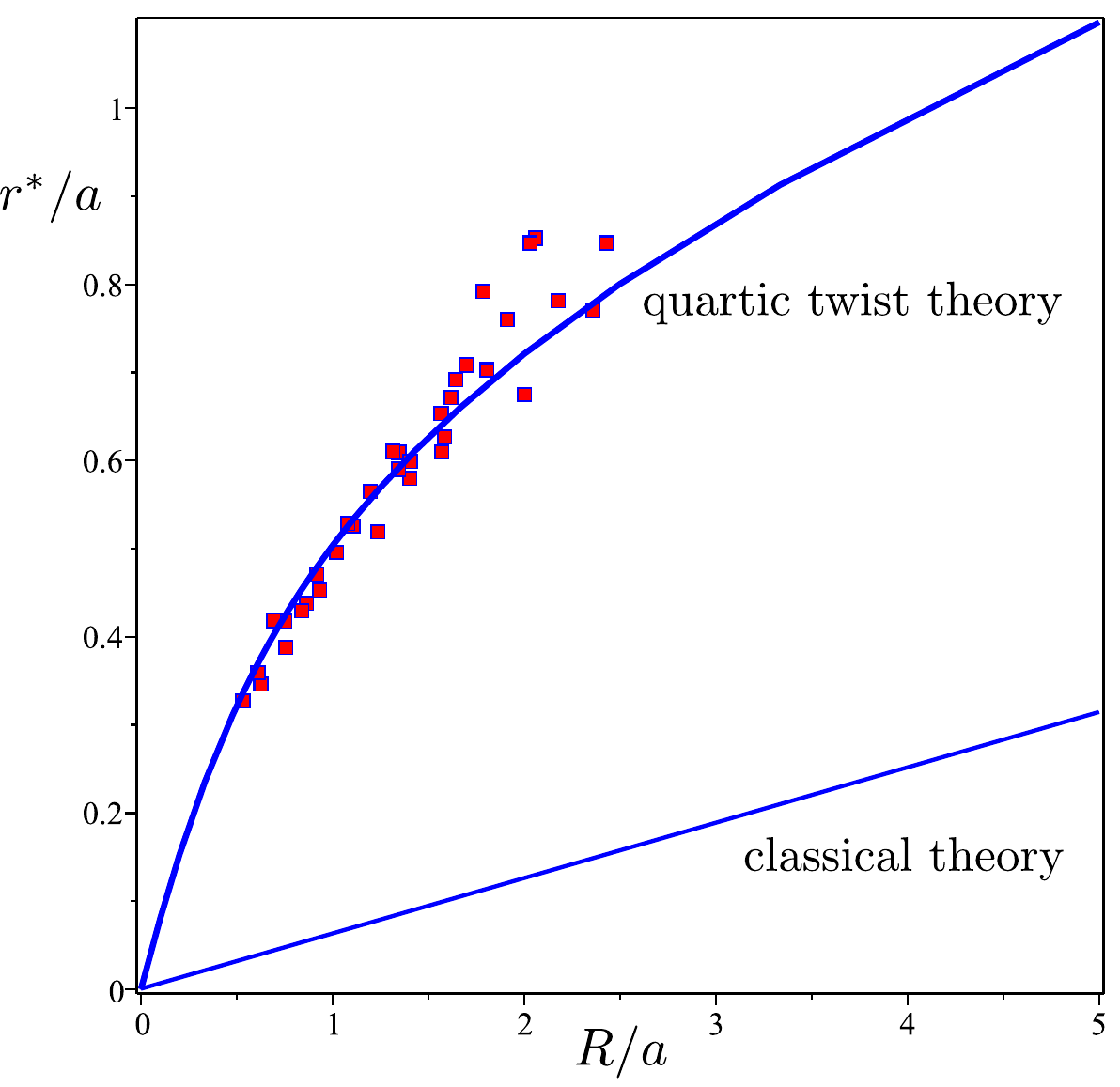}
		\caption{Experiment II; Observation 1.}
		\label{fig:universal_II_1}
	\end{subfigure}
	\begin{subfigure}[c]{0.30\linewidth}
		\centering
		\includegraphics[width=\linewidth]{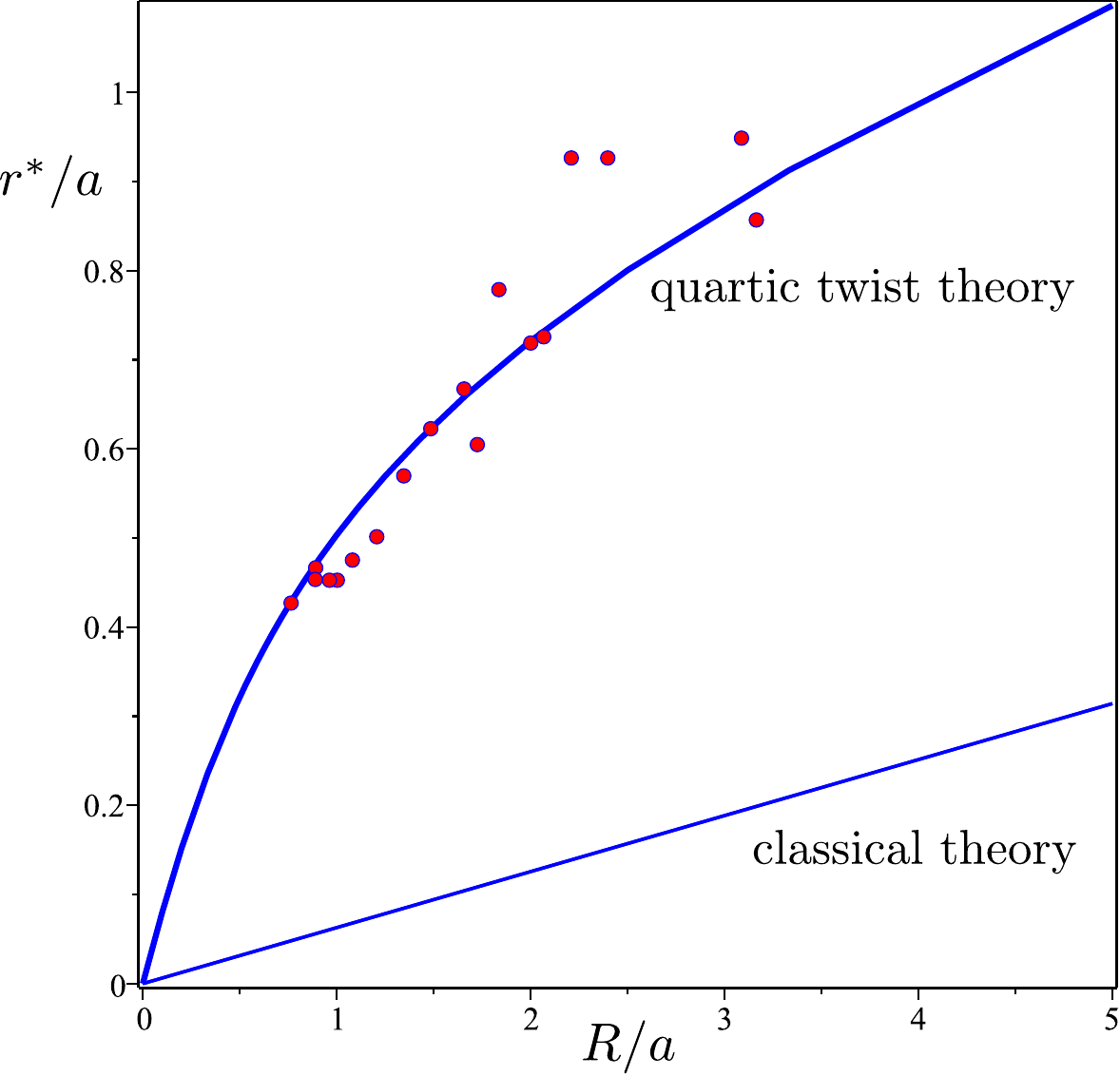}
		\caption{Experiment II; Observation 2.}
		\label{fig:universal_II_2}
	\end{subfigure}
	\caption{Disaggregation of the data shown in Fig.~\ref{fig:r_star_vs_R}. Experiment I uses a SSY solution at concentration $c=30\,\mathrm{wt}\%$ and temperature $\temp=25\,^\circ\mathrm{C}$; Experiments II uses a SSY solution at concentration $c=31.5\,\mathrm{wt}\%$ and temperature $\temp=25\,^\circ\mathrm{C}$. Observations are denoted $0$, $1$, or $2$, according to whether they took place with the cell just prepared, one, or two days later, respectively.}
	\label{fig:disaggregation_inversion_radius}
\end{figure}

Two distinct experiments were performed and, for each, three observations were made at different times. Experiments, conventionally denoted I and II, used different SSY solutions at room temperature. The data that were aggregated per experiment in Fig.~\ref{fig:r_star_vs_R} are now disaggregated in Fig.~\ref{fig:disaggregation_inversion_radius}.  

\section{Error Analysis and Statistical Dispersion}\label{sec:error_analysis}
The radii $R_i$ of microcavities and $r^\ast_i$ of inversion rings were measured on image reproductions; they were all affected by the same absolute error $\Delta R=\Delta r^\ast=1\,\mu\mathrm{m}$.
\begin{figure}
	\centering
	\begin{subfigure}[c]{0.3\linewidth}
		\centering
		\includegraphics[width=.95\linewidth]{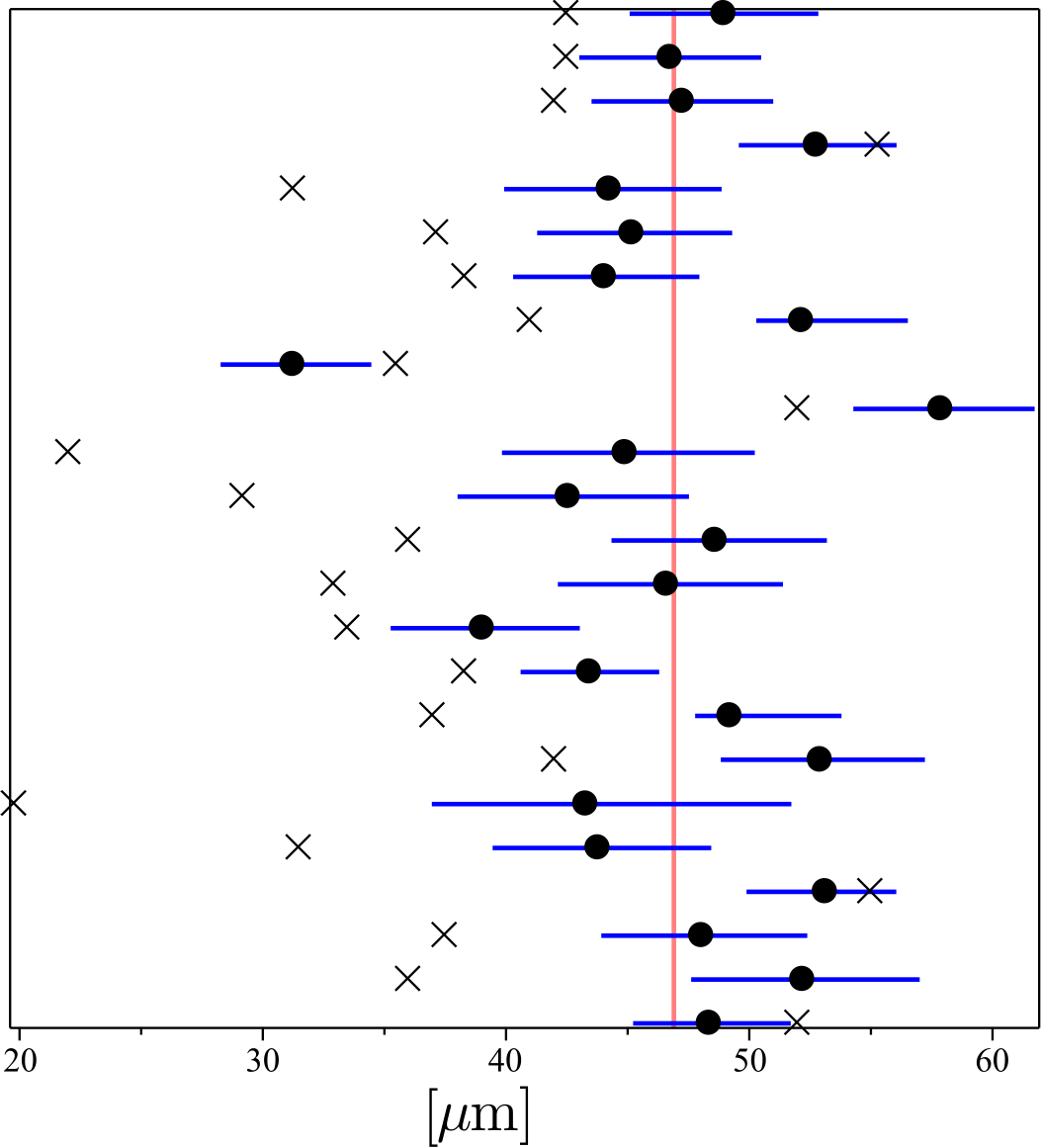}
		\caption{Experiment I; Observation 0: $\bar{a}\approx47\,\mu\mathrm{m}$.} 
		\label{fig:histogram_I_0}
	\end{subfigure}
	\begin{subfigure}[c]{0.3\linewidth}
		\centering
		\includegraphics[width=0.97\linewidth]{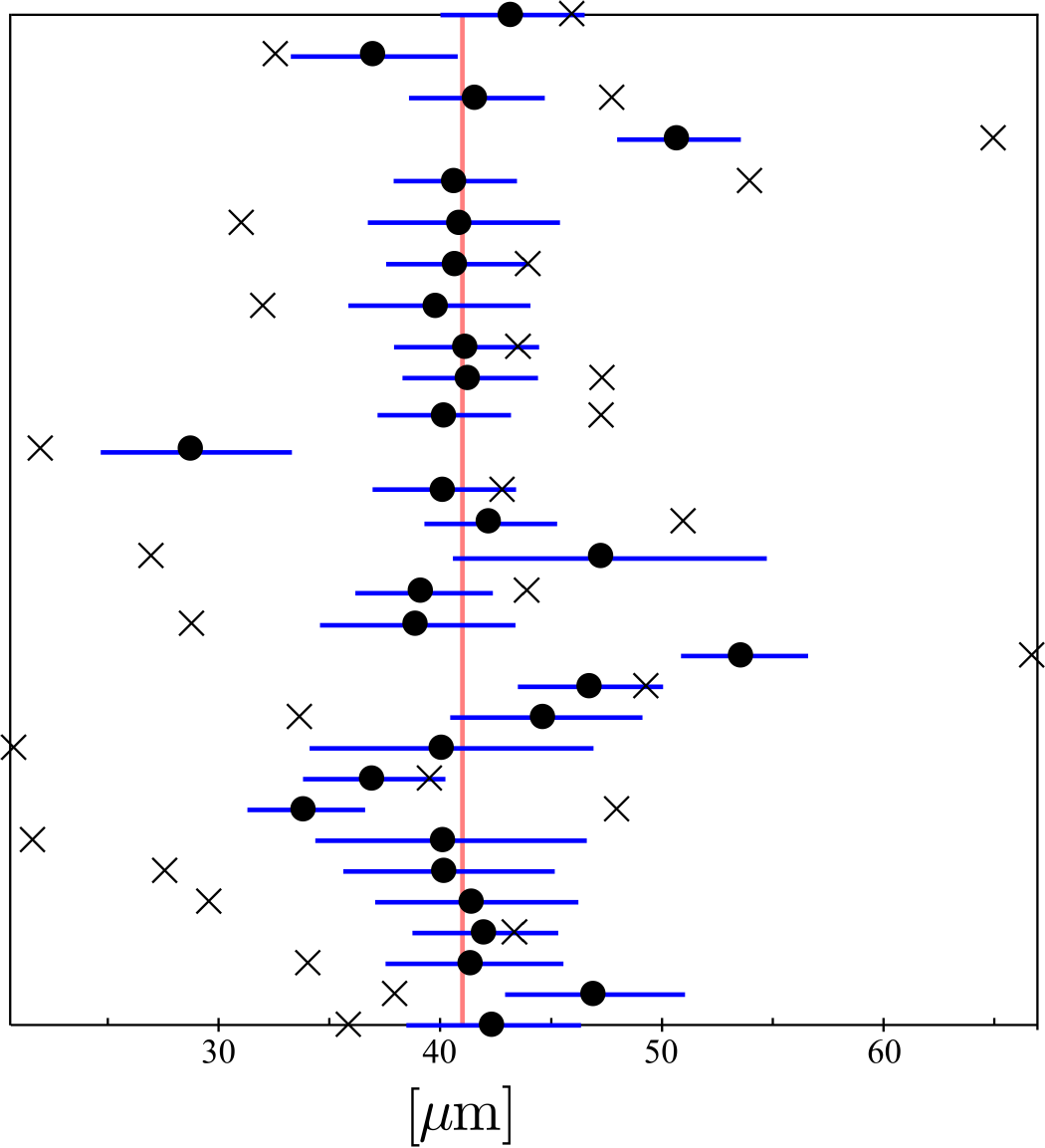}
		\caption{Experiment I; Observation 1: $\bar{a}\approx41\,\mu\mathrm{m}$.}
		\label{fig:histogram_I_1}
	\end{subfigure}
	\begin{subfigure}[c]{0.3\linewidth}
		\centering
		\includegraphics[width=0.95\linewidth]{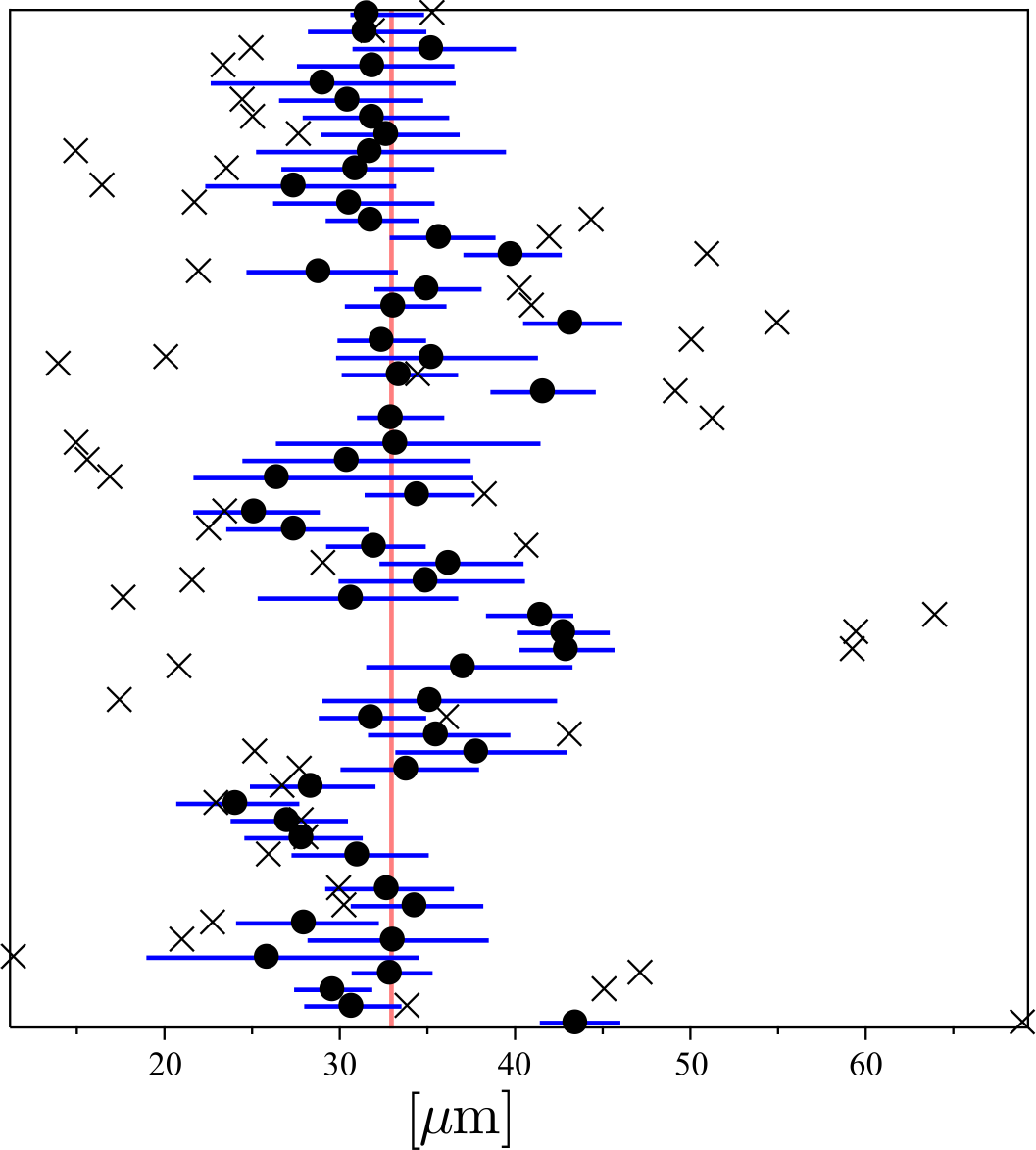}
		\caption{Experiment I; Observation 2: $\bar{a}\approx33\,\mu\mathrm{m}$.}
		\label{fig:histogram_I_2}
	\end{subfigure}
	\begin{subfigure}[c]{0.30\linewidth}
		\centering
		\includegraphics[width=.97\linewidth]{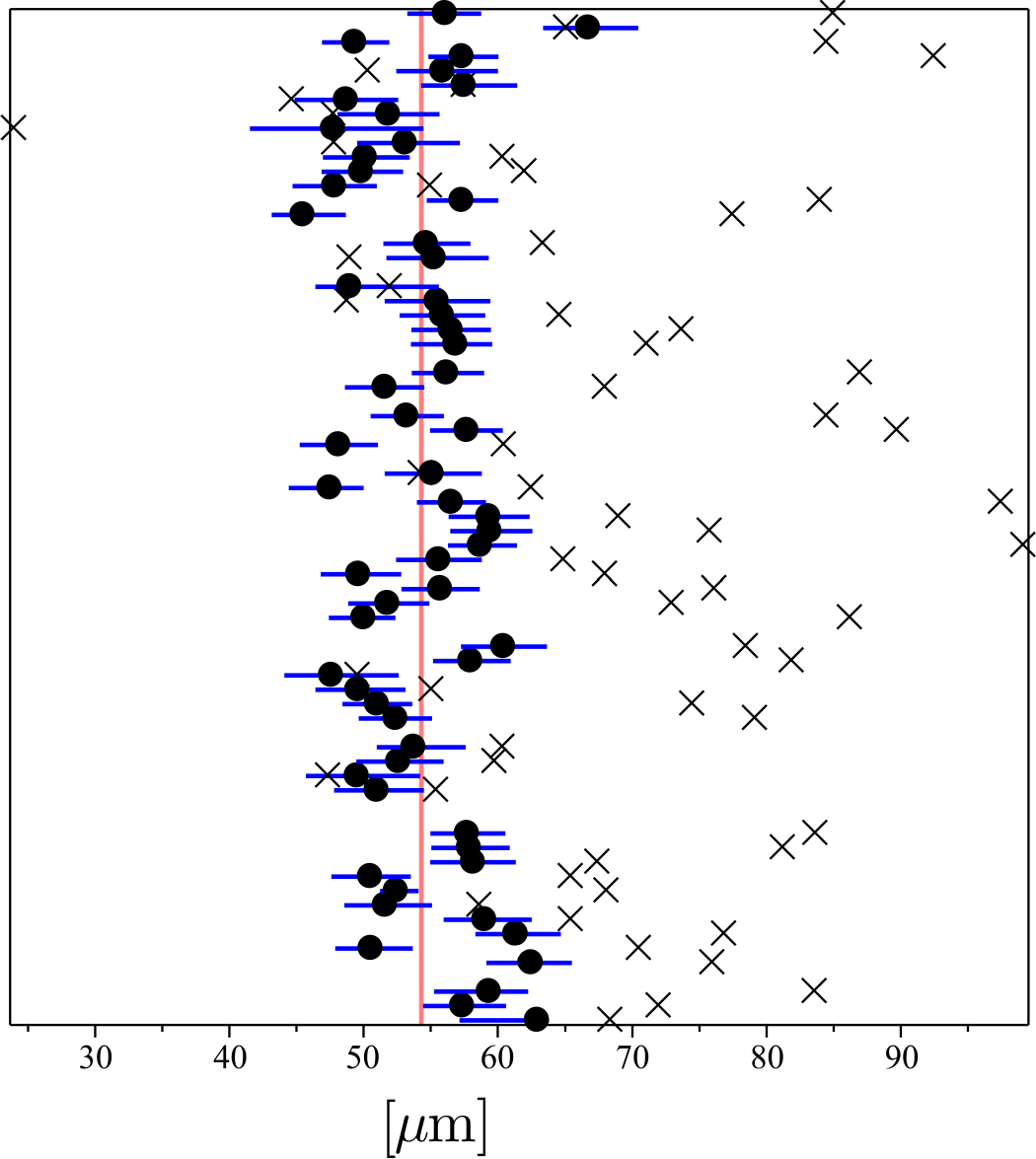}
		\caption{Experiment II; Observation 0: $\bar{a}\approx54\,\mu\mathrm{m}$.} 
		\label{fig:histogram_II_0}
	\end{subfigure}
	\begin{subfigure}[c]{0.3\linewidth}
		\centering
		\includegraphics[width=0.95\linewidth]{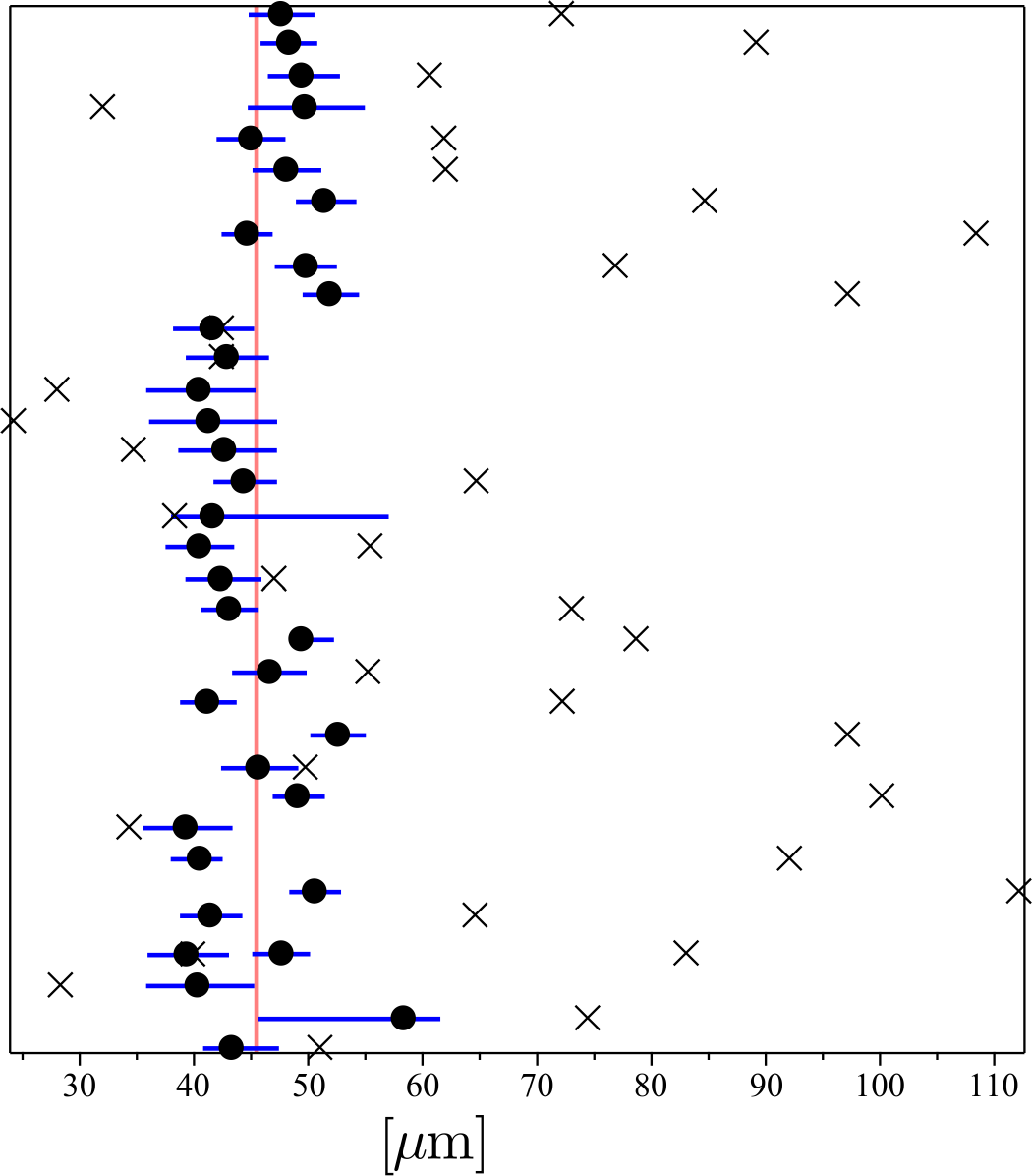}
		\caption{Experiment II; Observation 1: $\bar{a}\approx46\,\mu\mathrm{m}$.}
		\label{fig:histogram_II_1}
	\end{subfigure}
	\begin{subfigure}[c]{0.30\linewidth}
		\centering
		\includegraphics[width=0.92\linewidth]{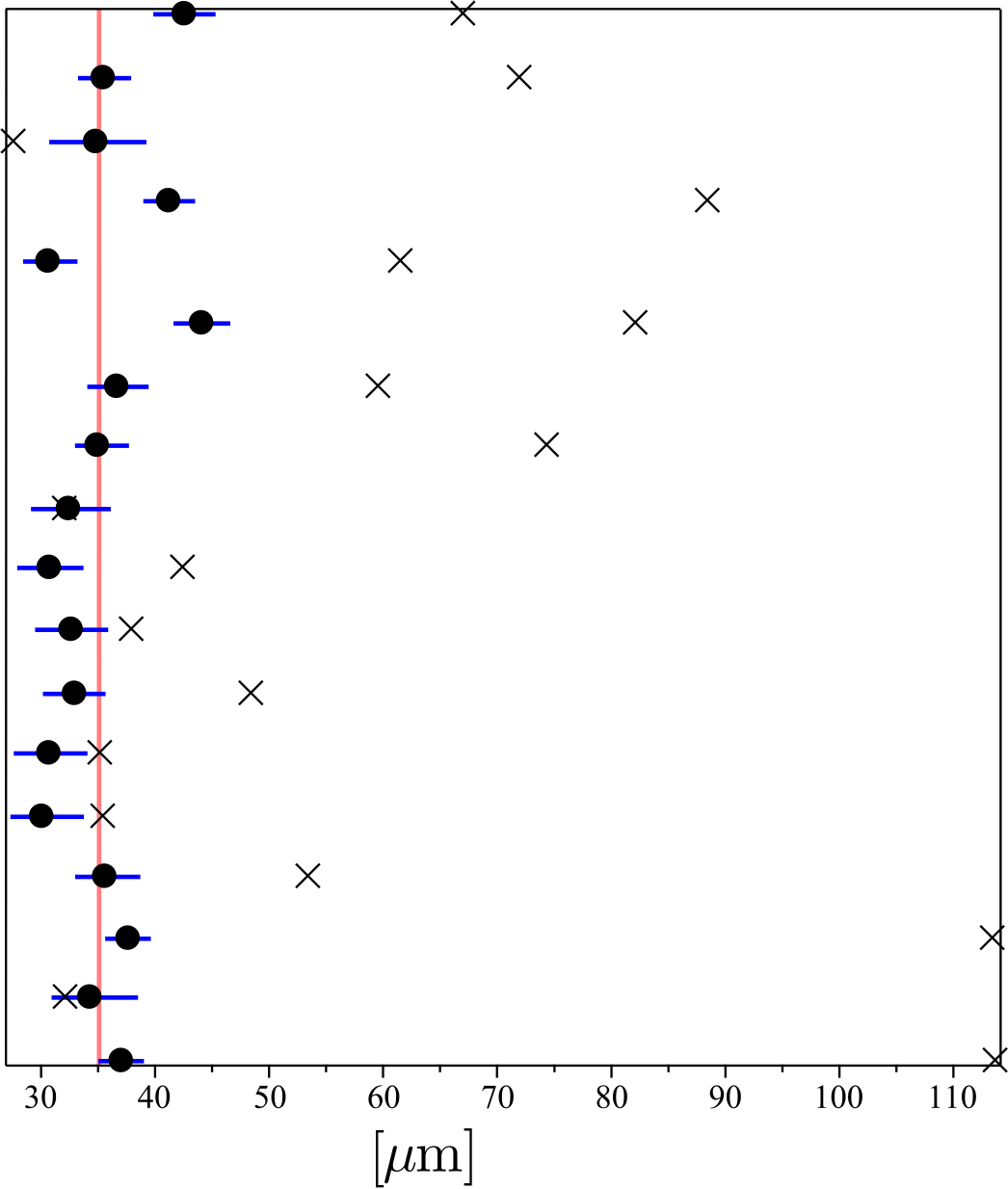}
		\caption{Experiment II; Observation 2: $\bar{a}\approx35\,\mu\mathrm{m}$.}
		\label{fig:histogram_II_2}
	\end{subfigure}
	\begin{subfigure}[c]{\linewidth}
		\centering
		\includegraphics[width=0.1\linewidth]{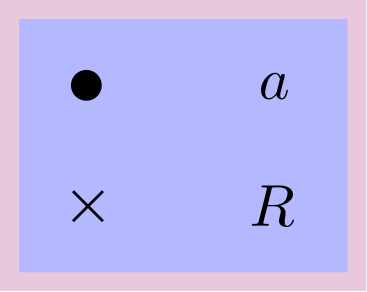}
	\end{subfigure}
	\caption{Dispersion histograms comparing the different measures inferred for $a$ to the cavity radii. They are organized according to experiments and observations, precisely as the panels of Fig.~\ref{fig:disaggregation_inversion_radius}; in addition, here $\bar{a}$ denotes the average value of $a$ for each set of conditions. All average values are  collected in Table~\ref{tab:a} in the main text.}
	\label{fig:disaggregation_a}
\end{figure}
As a consequence, the corresponding values of $\rho^\ast=r^\ast/R$ were listed as $\rho^\ast_i\pm\Delta\rho^\ast_i$, where
\begin{equation}
	\label{eq:rho_star_error}
	\Delta\rho^*_i:=\frac{r^*_{i}}{R_{i}}\left(\frac{1}{r^*_{i}}-\frac{1}{R_{i}}\right)\Delta r^*.
\end{equation}
Since, as shown in Fig.~\ref{fig:inversion_ring_radius}, the theoretical link between $\rho^\ast$ and $\lambda=a/R$ is not linear, the symmetric absolute errors $\pm\Delta \rho^\ast_i$ propagate asymmetrically to the corresponding $a_i$'s; we write them as $\pm\Delta a_i^\pm$, where
\begin{equation}
	\label{eq:Delta_a_i}
	\Delta a_i^\pm=R_i\Delta\lambda_i^\pm+\lambda_i\Delta R
\end{equation}
and $\Delta\lambda_i^\pm$ are defined by
\begin{equation}
\rho^\ast(\lambda_i\pm\Delta\lambda_i^\pm)=\rho^\ast_i\pm\Delta\rho^\ast_i.
\end{equation}
Thus, we can attribute to the $a_i$'s the following relative error,
\begin{equation}
	\label{eq:a_i_relative_error}
	\frac{\Delta a_i}{a_i}:=\frac12\left(\frac{\Delta a_i^+}{a_i}+\frac{\Delta a_i^-}{a_i}\right)=\frac{\Delta\lambda_i}{\lambda_i}+\frac{\Delta R}{R_i},
\end{equation}
where $\Delta\lambda_i:=(\lambda^+_i-\lambda^-_i)/2$.

The dispersion histograms for the $a_i$'s and $R_i$'s are shown in Fig.~\ref{fig:disaggregation_a} for every experiment and observation. 
Figure~\ref{fig:histogram_I_2} is the same as Fig.~\ref{fig:dispersion_histogram} in the main text; it is reproduced here for completeness.

To estimate the degree of statistical dispersion for both data sets $a_i$'s
and $R_i$'s, we computed the relative \emph{mean absolute deviation} $\mathcal{M}$, defined for a set of $n$ data $x_i$ by
\begin{equation}
	\label{eq:dispersion_mad}
	\mathcal{M}(x):=\frac{1}{\bar{x}}\frac{1}{n}\sum_{i=1}^n|x_i-\bar{x}|,
\end{equation}
where $\bar{x}$ denotes the (arithmetic) average of the $x_i$'s.
A summary of the averages $\bar{a}$ and $\bar{R}$, and the mean absolute deviations about them for all experiments and observations is presented in Table~\ref{tab:M_x}.
\begin{table}
	\begin{center}
		\begin{tabular}{|c|c|c|c|}
			\cellcolor{lightgray}{Experiment} & \cellcolor{lightgray}{Observation 0}  & \cellcolor{lightgray}{Observation 1} & \cellcolor{lightgray}{Observation 2}  \\ 
			\rule{0pt}{2ex}
			$\mathrm{I}$  &  $\bar{a}\approx 47 \, \mu\mathrm{m}, \quad \bar{R}\approx 38 \, \mu\mathrm{m}$ &  $\bar{a}\approx 41 \, \mu\mathrm{m}, \quad \bar{R}\approx 40 \, \mu\mathrm{m}$ & $\bar{a}\approx 33 \, \mu\mathrm{m}, \quad \bar{R}\approx 32 \, \mu\mathrm{m}$ \\
			\rule{0pt}{2ex}
			\quad & $\mathcal{M}(a)\approx8\%, \quad \mathcal{M}(R)\approx20\%$ & $\mathcal{M}(a)\approx8\%, \quad \mathcal{M}(R)\approx24\%$ & $\mathcal{M}(a)\approx9\%, \quad \mathcal{M}(R)\approx39\%$ \\
			\hline
			$\mathrm{II}$ & $\bar{a}\approx 54 \, \mu\mathrm{m}, \quad \bar{R}\approx 68 \, \mu\mathrm{m}$ & $\bar{a}\approx 46 \, \mu\mathrm{m}, \quad \bar{R}\approx 64 \, \mu\mathrm{m}$ &  $\bar{a}\approx 35 \, \mu\mathrm{m}, \quad \bar{R}\approx 60 \, \mu\mathrm{m}$ \\
			\rule{0pt}{2ex}
			\quad & $\mathcal{M}(a)\approx7\%, \quad \mathcal{M}(R)\approx17\%$ &  $\mathcal{M}(a)\approx9\%, \quad \mathcal{M}(R)\approx32\%$ &   $\mathcal{M}(a)\approx9\%, \quad \mathcal{M}(R)\approx36\%$ \\
			\hline
		\end{tabular}
	\end{center}
	\caption{Averages and corresponding relative mean absolute deviations for data sets $a_i$ and $R_i$ in every experiment and observation. The averages $\bar{a}$'s are the estimates for $a$ listed in Table~\ref{tab:a} in the main text.}
	\label{tab:M_x}
\end{table}
It is apparent from these data how the microcavity radii $R_i$ are much more scattered about their averages compared to the $a_i$. This justifies considering the averages $\bar{a}$'s as estimates of a phenomenological length characteristic only of the material and its physical conditions.

\input{Inversion_ring_in_SSY_spherical_droplets.bbl}
\end{document}

%% file: Inversion_ring_in_SSY_spherical_droplets.bbl
%